\documentclass[draftcls,onecolumn,12pt]{IEEEtran}
\pdfoutput=1
\ifCLASSINFOpdf
\else
\fi

\ifCLASSOPTIONcompsoc
\usepackage[caption=false,font=normalsize,labelfont=sf,textfont=sf]{subfig}
\else
\usepackage[caption=false,font=footnotesize]{subfig}
\fi
\usepackage{cite}
\usepackage{bm}
\usepackage{url}
\usepackage{hyphsubst, flushend}
\usepackage{amsmath}
\usepackage{mathtools}
\usepackage[]{amssymb}
\usepackage{amsthm,amsfonts}
\usepackage{float}
\usepackage{dsfont}
\usepackage{array}
\usepackage{booktabs}
\usepackage{multirow}
\usepackage{footmisc,caption}

\usepackage{graphicx}
\usepackage[para,flushleft]{threeparttable} 
\usepackage{dcolumn} 

\usepackage[ruled]{algorithm2e}
\usepackage{xifthen} 
\usepackage{enumitem}

\DeclareUnicodeCharacter{2061}{}

\makeatletter
\newcommand{\removelatexerror}{\let\@latex@error\@gobble}
\makeatother

\def \ba {\begin{array}}
\def \ea {\end{array}}
\def \benu {\begin{enumerate}}
\def \eenu {\end{enumerate}}
\def \bdes {\begin{description}}
\def \edes {\end{description}}

\def \bitem {\begin{itemize}}
\def \eitem {\end{itemize}}

\def \bfl {\begin{flushleft}}
\def \efl {\end{flushleft}}
\def \bfr {\begin{flushright}}
\def \efr {\end{flushright}}

\def \beq {\begin{equation}}
\def \eeq {\end{equation}}
\def \bqa {\begin{eqnarray}}
\def \eqa {\end{eqnarray}}
\def \bqa* {\begin{eqnarray*}}
\def \eqa* {\end{eqnarray*}}

\def \bal {\begin{align}}
\def \eal {\end{align}}

\newcounter{mytempeqncnt}

\DeclareMathOperator*{\argmax}{arg\,max}

\DeclareUnicodeCharacter{2061}{}

\begin{document}
%
\title{
Expectation Maximization Aided 
Modified Weighted Sequential Energy Detector 
for Distributed Cooperative Spectrum Sensing
}

\author{\IEEEauthorblockN{Mohammed Rashid,~\IEEEmembership{Member,~IEEE}, and 
Jeffrey A. Nanzer,~\IEEEmembership{Senior Member,~IEEE}}

\thanks{This work has been submitted to the IEEE for possible publication. Copyright may be transferred without notice, after which this version may no longer be accessible.}
\thanks{This effort was sponsored in whole or in part by the Central Intelligence Agency (CIA), through CIA Federal Labs. The U.S. Government is authorized to reproduce and distribute reprints for Governmental purposes notwithstanding any copyright notation thereon. The views and conclusions contained herein are those of the authors and should not be interpreted as necessarily representing the official policies or endorsements, either expressed or implied, of the Central Intelligence Agency. \textit{(Corresponding author: Jeffrey A. Nanzer.)}}
\thanks{The authors are with the Department of Electrical and Computer Engineering, Michigan State University, East Lansing, MI 48824 (e-mail: \mbox{rashidm@ieee.org,} nanzer@msu.edu).}}

\maketitle
\thispagestyle{empty}
\pagestyle{empty}

%
%


\def\bda{\mathbf{a}}
\def\bdd{\mathbf{d}}
\def\bde{\mathbf{e}}
\def\bdf{\mathbf{f}}
\def\bdg{\mathbf{g}} 
\def\bdh{\mathbf{h}}
\def\bdm{\mathbf{m}}
\def\bds{\mathbf{s}} 
\def\bdn{\mathbf{n}}
\def\bdu{\mathbf{u}}
\def\bdv{\mathbf{v}}
\def\bdw{\mathbf{w}} 
\def\bdx{\mathbf{x}} 
\def\bdy{\mathbf{y}} 
\def\bdz{\mathbf{z}}
\def\bdA{\mathbf{A}}
\def\bdC{\mathbf{C}}
\def\bdD{\mathbf{D}} 
\def\bdF{\mathbf{F}}
\def\bdG{\mathbf{G}} 
\def\bdH{\mathbf{H}}
\def\bdI{\mathbf{I}}
\def\bdJ{\mathbf{J}}
\def\bdU{\mathbf{U}}
\def\bdX{\mathbf{X}}
\def\bdK{\mathbf{K}}
\def\bdQ{\mathbf{Q}}
\def\bdR{\mathbf{R}}
\def\bdS{\mathbf{S}}
\def\bdV{\mathbf{V}}
\def\bdW{\mathbf{W}}
\def\bdGamma{\bm{\Gamma}}
\def\bdgamma{\bm{\gamma}}
\def\bdalpha{\bm{\alpha}}
\def\bdmu{\bm{\mu}}
\def\bdSigma{\bm{\Sigma}^n}
\def\bdOmega{\bm{\Omega}}
\def\bdxi{\bm{\xi}}
\def\bdl{\bm{\ell}}
\def\bdLambda{\bm{\Lambda}}
\def\bdeta{\bm{\eta}}
\def\bdPhi{\bm{\Phi}}
\def\bdpi{\bm{\pi}}
\def\bdtheta{\bm{\theta}}
\def\bdTheta{\bm{\Theta}}
\def\bddelta{\bm{\delta}}

\def\btau{\bm{\tau}}
\def\deg{\circ}

\def\tq{\tilde{q}}
\def\tbdJ{\tilde \bdJ}
\def\l{\ell}
\def\bdzero{\mathbf{0}} 
\def\bdone{\mathds{1}} 
\def\Exp{\mathbb{E}} 
\def\exp{\text{exp}} 
\def\ra{\rightarrow}

\def \Q{\mathcal{Q}}
\def\R{\mathbb{R}} 
\def\C{\mathbb{C}} 
\def\CN{\mathcal{CN}} 
\def\N{\mathcal{N}} 
\begin{abstract}
Energy detector (ED) is a popular choice for distributed cooperative 
spectrum sensing because it does not need to be cognizant of the 
primary user (PU) signal characteristics. 
However, the conventional ED-based sensing usually 
requires large number of observed samples per energy statistic, 
particularly at low signal-to-noise ratios (SNRs), 
for improved detection capability. 
This is due to the fact that it 
uses the energy only from the present sensing interval 
for the PU detection. 
Previous studies have shown that even with 
fewer observed samples per energy statistics, improved detection capabilities 
can be achieved 
by aggregating both present and past ED samples in a test statistic. 
Thus, a weighted sequential energy detector (WSED) has been proposed, but 
it is based on aggregating all the collected 
ED samples over an observation window. 
{For a highly dynamic PU over the consecutive sensing intervals, that involves also combining the outdated samples in 
the test statistic that do not correspond to the present state of the PU}. 
In this paper, we propose a modified WSED (mWSED) that 
uses the primary user states information over the window to aggregate only the 
highly correlated ED samples in its test statistic. In practice, 
since the PU states are a priori unknown, 
we also develop a joint 
expectation-maximization and Viterbi (EM-Viterbi) algorithm 
based scheme to iteratively estimate the states by 
using the ED samples collected over the window. 
The estimated states are then used in mWSED to compute its test statistics, 
and the algorithm is referred to here as the EM-mWSED algorithm. 
{Simulation results show that EM-mWSED outperforms other schemes} 
and its performance improves by increasing the average number of 
neighbors per SU in the network, 
and by increasing the SNR or the number of samples per energy statistic. 
\end{abstract}
		\vspace{-0.20in}
\section{Introduction}\label{Intro_section}
A cognitive radio system is an intelligent wireless communication system that learns from its surrounding radio environment and adapts its operating parameters (e.g., carrier frequency, transmit power, and digital modulation scheme) in real-time to the spatiotemporal variations of the RF spectrum. The primary objective of the cognitive scheme is to enable the unlicensed (secondary) users to opportunistically utilize the spectrum owned by the licensed (primary) users, where the reconfigurability of the radio is accomplished using software-defined radio based platforms 
\cite{Cognitive_survey, Cognitive_2, Haykin_2005}. 
{However, the opportunistic access of the wireless spectrum entails using 
the spectrum sensing algorithm at the secondary users (SUs) to 
detect the presence or absence of the primary user (PU) 
in the channel. For the detection of PU, several sensing algorithms have been presented over the past years including the energy detector (ED) \cite{ED_2007}, 
coherent ED \cite{Coherent_ED_2011}, matched filter detector \cite{MF_2018}, 
cyclostationary feature detector \cite{Cyclo_2014, Cyclo_2020}, information 
theoretic criterion based detector \cite{AIC_2010}, and eigen-values based 
detectors \cite{Eig_2017, Eig_2022}. These detectors 
make different trade-offs between the spectrum sensing delay, 
computational complexity, and the amount of PU's signal 
information needed for sensing. 
Among these detectors, ED stands out as a preferred choice because of its 
low computational complexity, ease of implementation, 
and due to the reason that it does not 
require any prior knowledge about the PU signal. 
As such, ED-based spectrum sensing has been exploited widely 
in the literature, e.g., \cite{KslotED_2022, ED_2023, IED_2012, Warit_2014}, 
including the present paper.}

Spectrum sensing (or PU detection) can be done by the SUs either by using a non-cooperative scheme or a cooperative scheme. In a non-cooperative scheme, each SU performs PU detection individually without any direct communication with the other SUs or a fusion center (FC) \cite{ED_2023, Eig_2022}. 
In contrast, in cooperative spectrum sensing, a group of SUs communicate with each other or with an FC to collaboratively perform the PU detection \cite{Warit_2014, Warit_2012, Kailkhura_2017}. Consequently, in comparison, the cooperative sensing approach is resilient to the deep fading and shadowing at an SU level, aids in eliminating the hidden terminal problem, reduces the sensing duration per SU, and demonstrates a better detection performance {across the SUs 
network}\cite{Ghasemi_2005, Cichon_2016}.

Cooperative spectrum sensing schemes can be further categorized into either a centralized scheme or a distributed scheme. In a centralized scheme, a fusion center collects the sensing information from the SUs, detects the unused band, and broadcasts the decision to the SUs \cite{Najimi_2013, Warit_2012, Warit_2014, 
Kaschel_2021}. However, the centralized approach is not scalable to large networks as the available communication resources are usually limited at 
the FC. Furthermore, an FC involvement defines a single point of failure for the centralized network. In comparison, in a distributed scheme, the SUs share their sensing statistics with their neighboring users followed by using a consensus protocol to collaboratively decide on the presence or absence of PU in the channel \cite{Li_2010, Mustafa_2021}. This approach not only eliminates the single point of failure from the network, but it is also scalable as the communication resources need to be shared only among the neighbors. 

The distributed cooperative spectrum sensing (DCSS) scheme usually has three critical phases, namely the sensing phase, the consensus phase, and the {channel use} phase. In the sensing phase, a group of SUs observes the PU channel for a certain time duration to collect a sufficient number of samples for computing the summary statistics (e.g., energy statistics \cite{Gharib_2020, ED_2023}). Next, in the consensus phase, the SUs share their summary statistics with their neighbors, 
and use an average consensus protocol \cite{Saber_2007_CP, Fast_MC_2004} to iteratively compute a weighted average of the globally shared values across the network. Upon consensus in such an approach, the final value is compared against a threshold at each SU to locally detect the presence or absence of 
the PU in the channel. Finally, in the 
{channel use} phase, the detection outcome is used to 
{decide on the opportunistic use of the PU channel} before restarting the cycle. This DCSS scheme was proposed in \cite{Li_2010} wherein the authors analyzed its convergence speed as well as the detection performance for varying false alarm rates. In \cite{Wang_2021, Qiao_2021}, DCSS was extended to protect against the eavesdropper attack by encrypting the summary statistics shared between the SUs, 
whereas in \cite{Kailkhura_2017, Moradi_2023}, 
the authors considered the scenarios in which some malicious SUs (aka Byzantines) 
may share falsified data with the neighbors and thus proposed algorithms 
to mitigate the Byzantine attacks on the network. 

The above-mentioned DCSS algorithms use the conventional approach in which each SU uses energy statistic only 
from the current sensing interval to make the PU 
detection. {This usually requires a large number of samples per 
energy statistic, specially for lower SNRs, in order to achieve improved PU detection. 
As a result, the sensing duration increases which in turn increases 
the power consumption per SU as well as reduces the network throughput 
\cite{Liu_2020}.
However, it has been shown that aggregating present and past ED samples 
at each SU, for lower SNRs, 
can result in an improved detection performance, even with using 
a fewer samples per energy statistics \cite{IED_2012, KslotED_2022, 
Warit_2012, Warit_2014}.} 
As such, in \cite{Warit_2014, Warit_2012}, 
a dynamic PU is modeled using a two-state Markov chain model and 
a weighted sequential energy detector (WSED) is proposed in which all the present and past ED samples over an observation window\footnote{An observation window is defined as a vector of 
length $D$ containing all the ED statistics from the 
$D-1$ past sensing periods as well as the ED statistic 
from the present sensing period.} are 
aggregated to achieve improved detection capability. 
However, \cite{Warit_2014, Warit_2012} assume a centralized scheme 
for cooperative spectrum sensing which as discussed before is not a scalable approach. 
{Using the average of past ED samples, 
a non-cooperative improved ED (IED) is also proposed in \cite{IED_2012} in which 
the averaged value compared against the threshold is used to improve the PU detection probability at lower SNRs. 
Furthermore, a non-cooperative multi-slot ED (mED) is developed in 
\cite{KslotED_2022} in which the ED samples from multiple sensing 
slots are compared against the threshold to detect the presence or 
absence of PU in the channel.}  

{In this paper, we consider a dynamic PU that follows a two-state Markov chain model for switching between the active and idle states \cite{Warit_2012, Warit_2014}. 
However, for the PU detection, we develop a test statistic for SUs wherein 
the highly correlated present and past ED samples are aggregated to 
improve the PU detection at lower SNRs with a few samples per ED. 
The SUs deploy the DCSS scheme in which the sensed test statistics are shared 
between the neighboring users to reach a consensus on the 
observed value across the network. 
The contributions made in this paper are listed as follows:} 
\begin{itemize} 
\item {A modified WSED (mWSED) is proposed in which only those 
past and present ED samples are aggregated in its test statistics 
that correspond to the PU's state in the current sensing interval. 
Thus, in contrast to the detectors presented in 
\cite{Warit_2014, KslotED_2022, IED_2012}, 
mWSED avoids including the outdated ED samples in its test statistics 
which improves the detection performance of SUs at lower SNRs (see the 
simulation results provided in Section 
\ref{sim_results_DCSS} and Section \ref{sim_results}).  
The closed-form equations for the probability of detection  
and probability of false alarm are also derived for mWSED.} 

\item {An underlying assumption in mWSED is that the actual present 
and past states visited by the PU are 
a priori known over the observation window. In practice, the states are 
unknown, and thus we develop a joint expectation-maximization (EM) and Viterbi 
based algorithm, referred to herein as EM-Viterbi, to estimate them 
using the ED samples collected over the window. 
Specifically, the EM algorithm provides an estimate of 
the model parameters of the joint probability distribution over the 
observation vector and the state vector. 
Next, using their estimate, the Viterbi algorithm \cite{Viterbi_1973} 
optimally estimates the latent state vector 
by the maximization and back tracing operations 
and using the properties of the two-state Markov chain model. 
The estimated state vector produced 
by the EM-Viterbi algorithm is then used in mWSED to aggregate only the 
highly correlated samples in its test statistic. The resulting algorithm is 
named here as the EM-mWSED algorithm. }
\item {Simulation results are included that demonstrate 
the estimation performance 
of EM-Viterbi and compare its detection performance with that of 
the EM-mWSED algorithm. 
Further, the detection performance of EM-mWSED is 
compared with the performances of other schemes in the literature. 
The results show 
that EM-mWSED performs better than the other methods 
and its detection performance improves by either 
increasing the average number of connections per SU in the network, or by 
increasing the SNR or the number of samples per energy statistics.}
\end{itemize}

This paper is outlined as follows. Section \ref{sys_mod} provides 
a brief review of the energy detection based spectrum 
sensing. 
Distributed cooperative spectrum sensing is discussed in Section \ref{DCSS_section}, 
also including a review of WSED and presentation 
of our proposed mWSED. Next, Section \ref{Section_EM} delivers an  
expectation-maximization and Viterbi algorithm based scheme for estimating 
the PU states over an observation window, 
including using it for mWSED. Simulation results are 
presented in Section \ref{sim_results}. Finally, Section \ref{conclusion_section} 
summarizes this work.

\section{Energy Detector Based Spectrum Sensing}\label{sys_mod}

We consider a distributed spectrum sensing system in which a network of 
$N$ SUs are spatially distributed and cooperating with each other to detect the PU in the channel.
As discussed in the previous section, we assume that the SUs deploy an energy based statistic to sense the channel. 
Thus, the energy computed by an $i$-th SU under the null hypothesis ($H_0$) and the alternate hypothesis ($H_1$) can be written as follows.
\begin{align}\label{ED_eqn}
x_i =\left\{
\begin{matrix}
\sum^L_{l=1}|n_{i,l}|^2, & \text{if  } H_0\\
\sum^L_{l=1}|h_i s_{l}+n_{i,l}|^2, & \text{if  } H_1\\
\end{matrix}\right.
\end{align}
in which $L$ is the total number of samples collected over the sensing interval, $h_i$ is the channel gain for SU $i$, $s_{l}$ represents 
the PU signal at time index $l$, and finally, $n_{i,l}$ denotes the  noise in the sensing interval which is assumed to be normally distributed with zero mean and variance $\sigma^2_n$. The signal to noise ratio (SNR) at the SU is defined by 
$\eta_i=\frac{\sum^L_{l=1}|h_{i} s_l|^2}{\sigma^2_n}$ which is $L$ times the SNR at the output of the energy detector.
{The ED statistics in \eqref{ED_eqn} 
follow a central chi-squared distribution with $L$ 
degrees of freedom under $H_0$, and a non-central chi-squared distribution 
with $L$ degrees of freedom and non-centrality parameter $\eta_i\sigma^2_n$ 
under $H_1$ \cite{proakis_book}. 
However, for the case of low SNR, we usually need a sufficient large number of 
samples ($L$) per energy statistics for improved PU detection\footnote{{This can 
be observed from Eqn. \eqref{ED_mV} that for smaller $\eta_i$ values 
we need larger $L$ values to realize non-overlapping distributions 
of ED statistics under both hypotheses and thereby 
achieve improved PU detection. However, note that larger $L$ implies larger 
sensing duration which increases the power consumption per SU as well as 
reduces the network throughput \cite{Liu_2020}. 
In this paper, we show that an improved detection 
capability can be achieved for lower SNRs and $L$ values by averaging the 
highly correlated present and past ED statistics.\label{ED_footnote}}}. Thus, 
we can invoke the central 
limit theorem (CLT) \cite{gubner2006} to assume that $x_i$ is Gaussian distributed under the hypothesis $H_h$ with mean $\mu_h$ and variance $\sigma^2_h$, for 
$h\in\{0,1\}$ \cite{KslotED_2022, IED_2012}. These means and variances can be computed as 
\begin{align}\label{ED_mV}
\mu_0 &= L\sigma^2_n\nonumber\\
\mu_1 &= (1+\eta_i)L\sigma^2_n\nonumber\\
\sigma^2_0 &= 2 L \sigma^4_n\nonumber\\
\sigma^2_1 &= 2(1+2\eta_i)L\sigma^4_n,
\end{align} 
As such, the probability of false alarm ($P_f$) and probability of detection 
($P_d$) can be 
computed as 
\begin{align}\label{ED_PdPfeqns}
P_f(\lambda) &= Q\left(\frac{\lambda-\mu_0}{\sigma_0}\right)\nonumber\\
P_d(\lambda) &= Q\left(\frac{\lambda-\mu_1}{\sigma_1}\right),
\end{align}
where $\lambda$ is the threshold for energy detection, and $Q(.)$ 
is the complementary cumulative distribution function of 
Gaussian distribution \cite{gubner2006}. Note that 
for a selected value of false alarm probability, the threshold $\lambda$ 
can be computed from \eqref{ED_PdPfeqns} by using the inverse of the $Q(.)$ function. 
Furthermore, given $\lambda$, the detection probability can be computed from 
\eqref{ED_PdPfeqns} as well. 
}

\begin{figure}[t]
	\centering
		\includegraphics[width=0.35\textwidth]{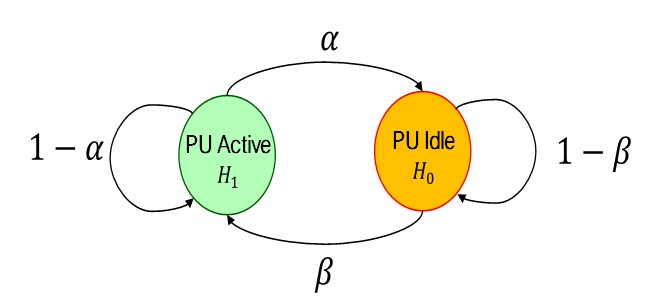}
	\caption{A graphical representation of the two-state Markov chain model describing the change in primary user activity over 
the sensing intervals. The parameters $\alpha$ and $\beta$ represent the transition probabilities of switching between the two states in the Markov model.}
	\label{fig:two_state}
\end{figure}
\begin{figure*}[tp]
    \begin{minipage}{0.33\textwidth}
     	\includegraphics[width=0.99\textwidth]{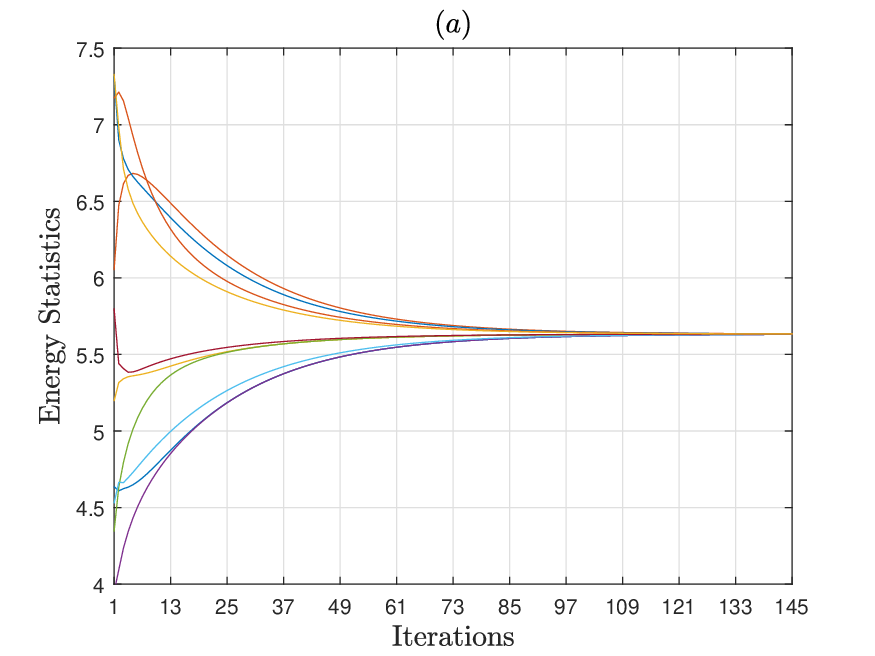}
    \end{minipage}\hspace{.01\linewidth}
    \begin{minipage}{0.33\textwidth}
\includegraphics[width=0.99\textwidth]{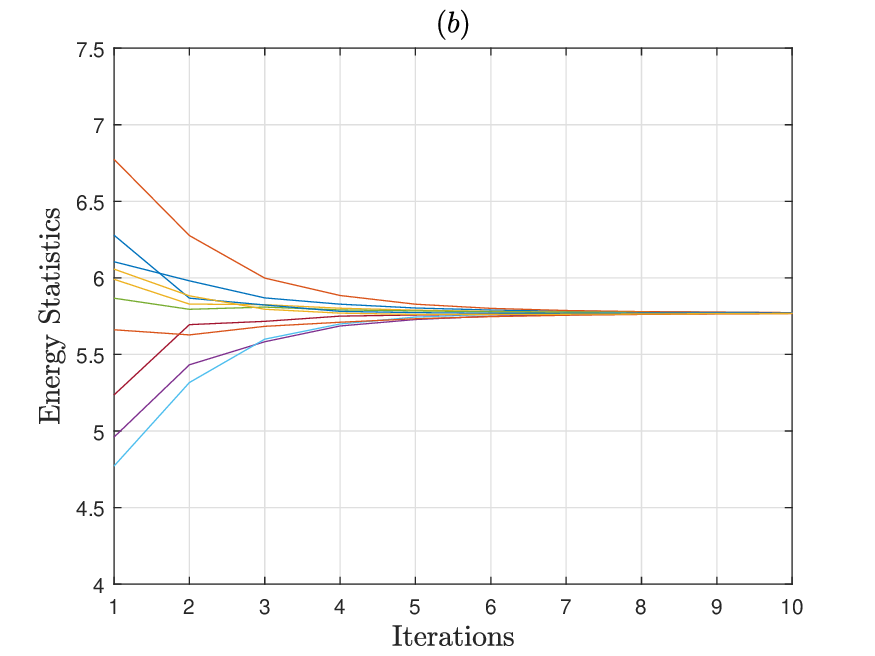}
		\end{minipage}
    \begin{minipage}{0.33\textwidth}
\includegraphics[width=0.99\textwidth]{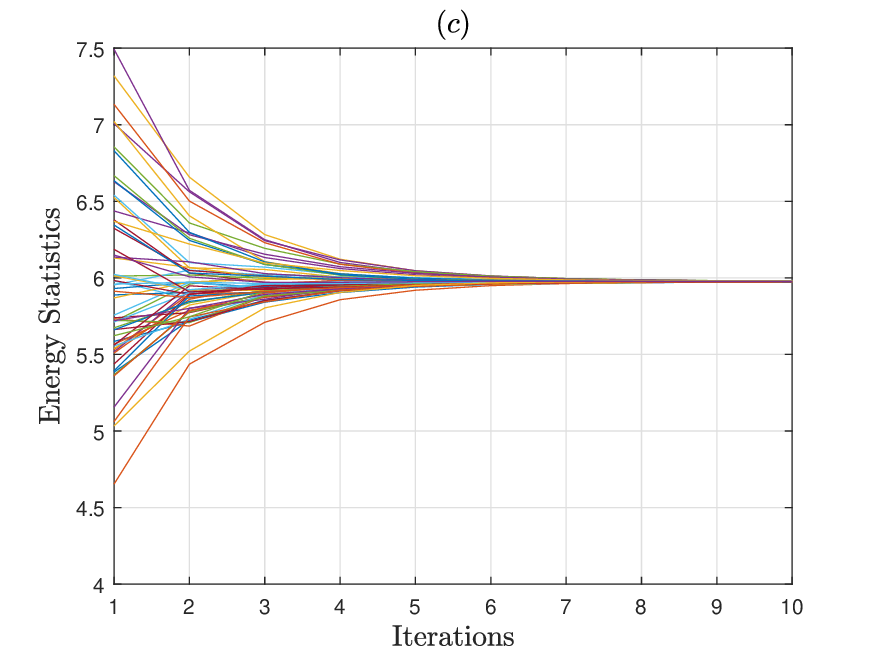}
		\end{minipage}
		\caption{Energy statistics of SUs from a single trial 
		vs. DCSS iterations for $(a)$ $N=10$, $c=0.2$, 
		$(b)$ $N=10$, $c=0.5$, and $(c)$ $N=60$, $c=0.2$, 
		when SNR $=-3$ dB, number of samples per energy statistic 
		$L=12$, and the PU follows the two-state Markov model with $\alpha=\beta=0.1$.}
				\label{fig_1:consensus}
\end{figure*}
\begin{figure*}[tp]
    \begin{minipage}{0.33\textwidth}
     	\includegraphics[width=0.99\textwidth]{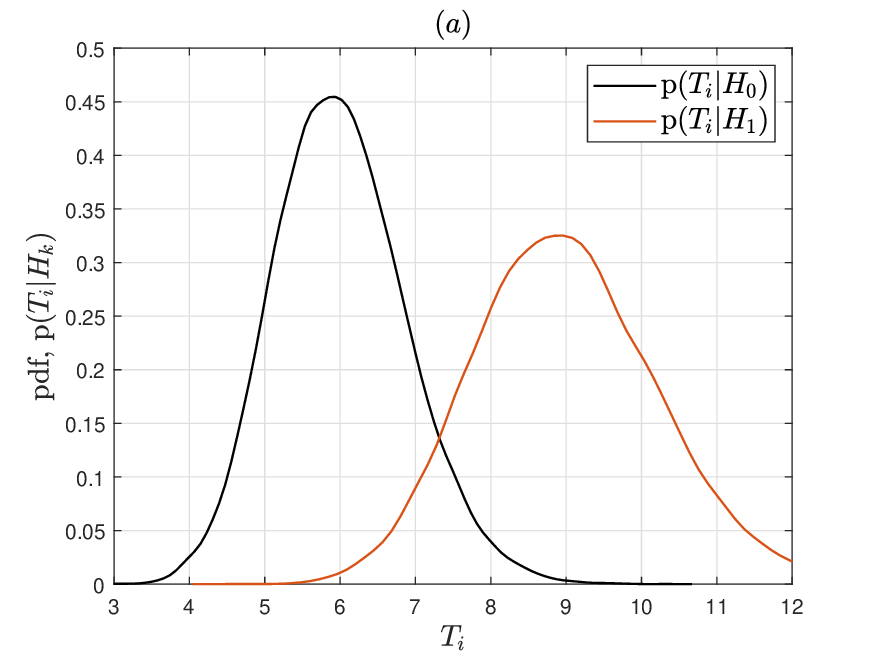}
    \end{minipage}\hspace{.01\linewidth}
    \begin{minipage}{0.33\textwidth}
\includegraphics[width=0.99\textwidth]{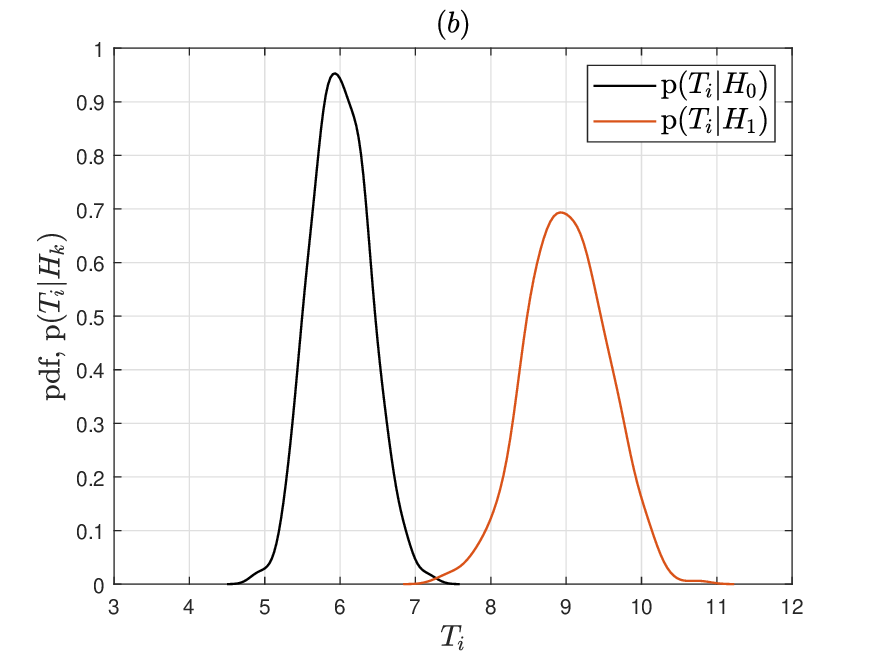}
		\end{minipage}
    \begin{minipage}{0.33\textwidth}
\includegraphics[width=0.99\textwidth]{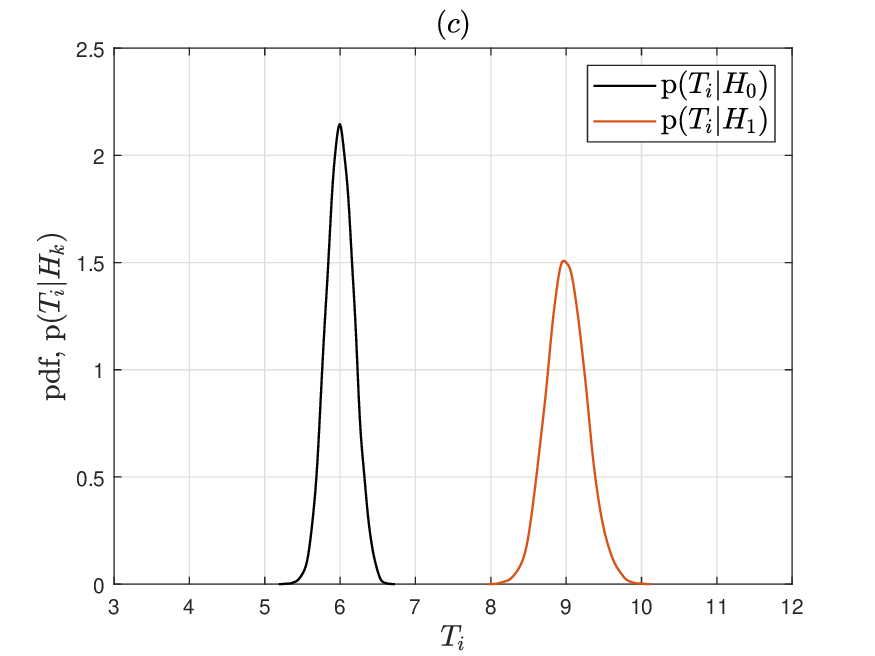}
		\end{minipage}
		\caption{{Probability density functions (pdfs) of 
		mWSED test statistics $T_i$ for SU $i$ under $H_0$ and $H_1$ with uniformly 
		distributed non-zero weights 
		when the number of averaged ED samples are 
		$(a)$ $C=4$, $(b)$ $C=20$, and $(c)$ $C=90$, 
		with SNR $=-3$ dB and $L=12$.}}
				\label{fig_mWSED1:distributions}
\end{figure*}

\section{Distributed Cooperative Spectrum Sensing}\label{DCSS_section}

{We consider a scenario wherein the PU follows a two-state Markov chain model for switching between the $H_0$ and $H_1$ as shown in Fig. \ref{fig:two_state} 
\cite{Warit_2012, Warit_2014}. 
Specifically, the next state visited by the PU depends only on its immediate previous state. Accordingly, in this figure, 
the parameter $\alpha$ denotes the transition probability of switching to an idle state ($H_0$) given that previously the PU was in the active state ($H_1$), whereas 
$\beta$ represents the transition probability of switching to an 
active state ($H_1$) if previously the PU was in the idle state ($H_0$). 
Thus, higher values of $\alpha$ and $\beta$ imply a highly dynamic PU, whereas their  smaller values represent its slowly time-varying behavior. 
In the following, we first describe a scalable 
distributed cooperative spectrum sensing scheme as considered here, 
followed with a brief review of the WSED algorithm 
and our modified WSED algorithm. }

Consider a network of SUs represented by an undirected graph $\mathcal{G}=(\mathcal{V},\mathcal{E})$ 
in which $\mathcal{V}=\{1,2,\ldots, N\}$ is the set of $N$ number of SUs in the network, 
and $\mathcal{E}=\{(i,j), \forall i,j\in V\}$ represents the set of all possible bidirectional 
communication links between them. 
The connectivity of the network in any realization is denoted by $c$ which is defined as a ratio of the number of active connections in the network ($N_a$) to the number of all possible connections among the SUs ($N(N-1)/2$). 
We consider a strongly connected network in which the neighboring 
users that are one hop away from each other share their information to reach 
a consensus. 
Therefore, wireless and 
computing resources such as bandwidth, processing power, and data storage capabilities, need to be only locally managed at the SUs, and scale 
proportionally to the average number of connections per SU in a network. 

A distributed cooperative spectrum sensing algorithm is a scalable and a fully distributed approach which deploys a consensus protocol at an SU. The 
protocol iteratively updates the sensing information at the user, by using the 
locally shared information, to reach consensus with the other users in the network. 
To elaborate, let $y_i (0)=x_i$ represents the initial energy statistic for an $i$-th SU, then in iteration $k$ of the DCSS algorithm, the SU updates 
its estimate by using a weighted average method as follows.
\begin{equation}\label{DCSS_eqn}
y_i (k)=y_i (k-1)+\sum_{j\in N_i} w_{ij} (y_j (k-1)-y_i (k-1))
\end{equation}
in which $N_i$ is the set of neighboring users of the $i$-th SU. The weight $w_{ij}$ can be selected as 
$w_{ij}=1/\text{max}⁡(d_i,d_j)$  
where $d_i$ and $d_j$ represent the number of neighbors of SU $i$ and SU $j$, respectively. This selection of weights results in a doubly-stochastic Metropolis-Hasting weighting matrix which guarantees convergence of the consensus algorithm \cite{Fast_MC_2004}. Thus, starting with a set of initial values $\{y_i (0), \text{ for } 
i=1,2,\ldots, N\}$, the algorithm running 
locally at each SU iteratively updates the values using \eqref{DCSS_eqn} until it converges to an average of the globally shared values across the network. The 
average value is defined by $y^*=\frac{\sum_{i=1}^N x_i}{N}$. 
Upon convergence, the decision can be made locally at the $i$-th SU by using the following rule
\begin{equation}\label{decision_rule}
d_i=\left\{
\begin{matrix}
H_1 & \text{if  } y^*\geq\lambda \\
H_0 & \text{otherwise}
\end{matrix}\right.
\end{equation}

\subsection{Simulation Results}\label{sim_results_DCSS}
Herein, we analyze the consensus 
performance of an energy detector (ED) based DCSS scheme when 
the network of SUs is randomly generated for different number of SUs $N$ and 
with varying connectivity $c$. The primary user follows a two-state Markov 
chain model for switching states between $H_0$ and $H_1$ over the multiple 
sensing intervals. Each SU uses $L=12$ samples for computing the 
energy statistic following \eqref{ED_eqn}, and has an SNR$=-3$ dB for the PU 
channel. In Fig. \ref{fig_1:consensus}, we demonstrate the convergence 
performance of the ED-based DCSS algorithm from a single trial 
for $N=10$ and $60$ users when 
the connectivity is either $c=0.2$ or $0.5$. It is observed that when 
$N=10$ and $c=0.2$, the consensus occurs in about $96$ iterations, but 
when the connectivity increases to $c=0.5$, it happens in about $7$ 
iterations. Similar observation is made if $N$ increases from $10$ to $60$ SUs 
for $c=0.2$. This is because the average number of connections per SU, i.e., 
$R=c(N-1)$, increases when either $c$ or $N$ increases, and thus 
the local averages computed at the SUs in \eqref{DCSS_eqn} are more 
accurate and stable resulting in the faster convergence speed.

Next, we first briefly review the WSED detector of 
\cite{Warit_2014} and discuss its extension for using it with the DCSS algorithm in \eqref{DCSS_eqn}. 
After that, we propose a modified WSED which as shown through simulation results
outperforms the DCSS-based WSED algorithm.
\subsection{Weighted Sequential Energy Detector}\label{WSED_section}

As proposed in \cite{Warit_2012, Warit_2014}, the WSED algorithm computes 
a weighted sum of all the present and past ED samples over an observation 
window of length $D$ to define a new 
test statistic, which for the $i$-th SU 
is given by 
\begin{equation} \label{WSED_eqn}
S_i=\sum_{d=1}^{D} w_d x_{i,d},
\end{equation}
where $x_{i,d}$ represents the energy statistic of the SU $i$ (as in \eqref{ED_eqn}) in the sensing interval $d$, with $x_{i,D}$ representing the energy 
at the present sensing interval. Thus, a total number of $D$ 
present and past ED samples are combined in the 
WSED statistic. The weights obey $\sum_{d=1}^D w_d =1$ 
and the authors in \cite{Warit_2012, Warit_2014} proposed to use 
exponential weights ($w_d=e^d/\sum_{d=1}^D e^d$) for an improved detection 
capability. Specifically, 
the exponential weighting is motivated to reduce the impact of aggregating 
the outdated past samples in \eqref{WSED_eqn} in a highly dynamic scenario. 
However, a centralized scheme is considered in \cite{Warit_2012, Warit_2014} wherein the SUs forward their statistics in \eqref{WSED_eqn} to 
a fusion center where a decision is made using an OR rule. 
As discussed before in Section \ref{Intro_section}, 
the use of a fusion center is a non-scalable approach, so 
in Section \ref{sim_results}, we extend WSED to the DCSS scheme 
of \eqref{DCSS_eqn} with consensus averaging 
performed on the ED samples aggregated in 
\eqref{WSED_eqn}. 
Upon consensus, the decision can be made locally at each SU by comparing $S_i$ against a threshold. 
Finally, as pointed out in \cite{Warit_2014}, the exact closed-form expressions for the probability of detection and the probability of false alarm for WSED are in general intractable to compute analytically, due to the aggregation of ED samples that may correspond to different states of the PU. However, the authors in \cite{Warit_2014} have derived approximated expressions which are also applicable for the DCSS scheme based WSED. 

\subsection{Modified Weighted Sequential Energy Detector}\label{mWSED_section}

\begin{figure}[t]
	\centering
		\includegraphics[width=0.35\textwidth]{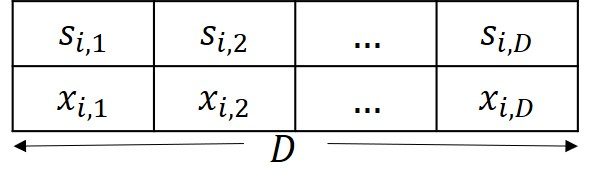}
	\caption{A notional view of the correspondence between the primary user states 
	$\left(\{s_{i,d}\in\{0,1\}, \forall d=1,2,\ldots,D\}\right)$ and the ED samples 
	$\left(\{x_{i,d}, \forall d=1,2,\ldots,D\}\right)$ 
	collected by the SU $i$ in an observation window of length $D$.}
	\label{fig:obser_window}
\end{figure}

In this subsection, we present our proposed modified WSED (mWSED). 
It is based on the motivation that instead of 
combining all the present and past ED samples over 
the observation window of length $D$, 
we combine only those ED samples in the summary statistic upon consensus that belong to the present state of the PU. 
As such, in mWSED, we begin by assuming that the states visited by the PU over the observation window are known to each SU. Notably, this 
assumption provides a starting point to derive mWSED, but later 
on in Section \ref{Section_EM} we also 
develop the EM-Viterbi algorithm to compute an estimate of those states at each SU, using the ED samples collected over the window as shown in 
Fig. \ref{fig:obser_window}. Thus, by comparing the present and past states over the observation window, each SU locally combines only 
those samples that correspond to the state of the PU in the present sensing interval. 
Therefore, the test statistic computed by mWSED at the $i$-th SU in 
the present $D$-th sensing interval is defined as 
\begin{equation}\label{mWSED_eqn}
T_i =\sum_{d=1}^{D} x_{i,d}\bdone\left(s_{i,d}=s_{i,D}, w_d\right),
\end{equation}
where $x_{i,d}$ represents the energy computed by SU $i$ in the $d$-th sensing interval and $s_{i,d}$ is the PU's state in that interval with 
$s_{i,d}=0$ denoting $H_0$ and $s_{i,d}=1$ implying $H_1$. 
$\bdone(A, w_d)$ is a weighted 
indicator function which outputs a non-zero weight $w_d$ for aggregating $x_{i,d}$ 
if $A$ is true, and outputs $w_d=0$ if $A$ is false. Thus, given the state information of PU over the observation window, either $x_{i,d}$ is included or excluded from $T_i$. Notably we assume here that the non-zeros 
weights are normalized and as such sum to $1$. 
%
Finally, the statistic $T_i$ is compared 
against a threshold at SU $i$ to make the PU detection.

{Now, let $C$ be the number of samples with the 
non-zero weights in \eqref{mWSED_eqn}, 
we note that for the assumed perfect PU's states information in mWSED, 
the use of uniformly distributed weights (i.e., $1/C$)
is optimal that do not change the means but 
aids in reducing the variance of $T_i$, with increasing $C$, 
under both $H_0$ and $H_1$ (see Propositions $1$ and $2$ in \cite{Warit_2014} for 
proofs). 
Furthermore, for this case, 
the test statistic $T_i$ is central chi-squared distributed with $CL$ degrees of 
freedom under $H_0$, and it is non-central chi-squared distributed with $CL$ 
degrees of freedom under $H_1$ and non-centrality parameter $C\eta_i\sigma^2_n$. 
In both these cases, we can define the threshold as $C\lambda$ to 
obtain the chi-squared distributions. However, for lower SNR and adequate 
number of samples per ED, we can use the CLT assumption \cite{gubner2006} 
as before to assume that $T_i$ is Gaussian distributed under $H_h$ with 
mean $m_h$ and variance $v^2_h$ for $h\in\{0,1\}$. These means and variances 
are given by 
\begin{align}\label{mWSED_mV_Eqns}
m_h &= \Exp[T_i|H_h]= \mu_h\sum^C_{c=1} w_c\nonumber\\
v^2_h &=\text{Var}[T_i|H_h]= \sigma^2_h\sum^C_{c=1} w^2_c,
\end{align}
where $\mu_h$ and $\sigma^2_h$ are defined in \eqref{ED_mV} for $h\in\{0,1\}$. 
The summation with index $c$ is used here to simplify \eqref{mWSED_eqn} and 
average the non-zero weighted ED samples 
where the weight for the $c$-th ED sample is defined by $w_c$. 
Since the normalized weights are assumed herein, 
we can note that the means are not effected by the choice of weights 
in \eqref{mWSED_mV_Eqns} 
but using $w_c=1/C$ aids in reducing the variances of 
$T_i$ under both $H_0$ and $H_1$ with increasing $C$. Also, note that 
larger values of $C$ can be realized for mWSED by equivalently 
increasing the length of the observation window ($D$) which enables the 
SUs to combine large number of present and past ED samples. However, 
this in turn results 
in improving the detection performance of mWSED as shown later 
in Section \ref{sim_mWSED}. 
Finally, using \eqref{mWSED_mV_Eqns}, the probability of false alarm 
and probability of detection can be given by 
\begin{align}\label{mWSED_PdPf_Eqns}
P_f(\lambda) &= Q\left(\frac{\lambda-m_0}{v_0}\right)\nonumber\\
P_d(\lambda) &= Q\left(\frac{\lambda-m_1}{v_1}\right),
\end{align}
where for a chosen 
false alarm probability, the threshold $\lambda$ can be computed from 
\eqref{mWSED_PdPf_Eqns} 
by using the inverse $Q(.)$ function. Together with the 
resulting detection probability, the 
operating point for an SU can be defined as $(P_f (\lambda),P_d (\lambda))$.}
\begin{figure}[t]
	\centering
		\includegraphics[width=0.4\textwidth]{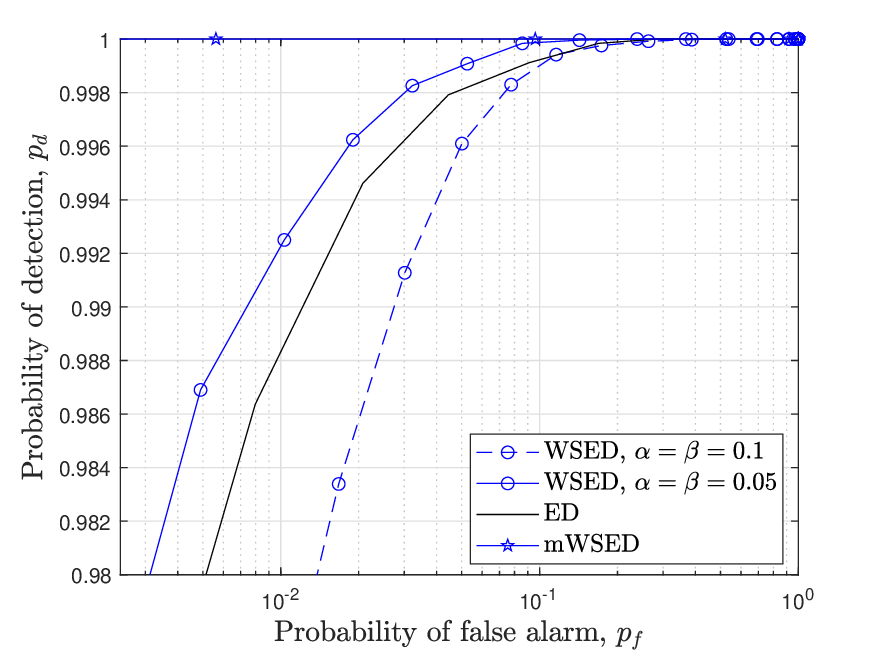}
	\caption{{Receiver operating characteristic curves for mWSED, WSED, and 
	conventional ED when $N=10$, $c=0.2$, $L=12$, SNR$=-3$ dB, 
	and PU follows a two-state Markov chain with varying $\alpha$ and $\beta$.}}
	\label{fig:ROC_mWSED}
\end{figure}

\subsection{Simulation Results}\label{sim_mWSED}
{In this subsection, we demonstrate the detection performance of 
mWSED using the simulation setup of Section \ref{sim_results_DCSS}. Thus, 
we assume a randomly generated network of $N=10$ SUs with 
connectivity $c=0.2$, and with $L=12$ samples per ED and SNR$=-3$ dB.} 

{In Fig. \ref{fig_mWSED1:distributions}, 
we first show the probability density functions (pdfs) of mWSED's test 
statistic $T_i$ for SU $i$, under both $H_0$ and $H_1$, 
when varying number of ED samples 
$C$ are averaged in $T_i$. As discussed above for Eqn. \eqref{mWSED_mV_Eqns}, 
it is observed in this figure 
that the means of the test statistic $T_i$ do not change 
but its variances under both hypotheses decrease with increasing $C$. 
This elucidates that the detection performance of 
mWSED will improve when large number of highly correlated ED samples are 
averaged in its test statistics.}

{Since a larger value of $C$ can be achieved for mWSED by choosing a longer observation window, in Fig. \ref{fig:ROC_mWSED}, we consider the window length as 
$D=150$, for the same network configuration as above, and demonstrate the 
receiver operating characteristic (ROC) curves for mWSED, WSED, and the 
conventional ED. The PU follows a two-state Markov model for 
switching between the active and idle states, and we consider 
both a slowly time varying PU with $\alpha=\beta=0.05$ and a 
highly dynamic PU with $\alpha=\beta=0.1$. We observe that the ED's 
performance does not vary with 
$\alpha$ and $\beta$ as it only considers the sample corresponding 
to the present state of the PU. 
It is observed that the WSED scheme shows poorer performance 
for the highly dynamic PU but outperforms ED for a slowly time-varying PU. 
This is because there are more chances of averaging the outdated samples in WSED 
in the former case than in the latter one. 
In contrast, since our mWSED method uses the PU states information 
to aggregate only the highly correlated ED samples in its test statistics, it 
outperforms both WSED and ED, and its 
performance is independent of the time-varying nature of PU.}

\section{Expectation Maximization Based State 
Estimation for Dynamic Primary User}\label{Section_EM}

The mWSED algorithm described in the previous section assumes that the actual 
states visited by the PU over the observation window are a priori known to the SUs. 
In practice, this may not be a valid assumption, and thus in this section we aim to 
compute an estimate of the states locally at each SU from the samples 
collected over the observation window.
 
To begin, let an SU $i$ collect $D$ ED samples over consecutive sensing intervals using \eqref{ED_eqn}, 
denoted by $\bdx_i=[x_{i,1},x_{i,2},\ldots,x_{i,D}]^T$ with $T$ representing 
the transpose operation. Using the CLT assumption 
\cite{gubner2006, Kailkhura_2017}, 
we assume that $x_{id}$ follows a Gaussian distribution represented by 
$\N\left(x_{i,d}|\mu_h,\sigma^2_h\right)$ with mean $\mu_h$ and variance $\sigma^2_h$, and with $h=0$ when the PU is idle, and $h=1$ when the PU is active. 
These means and variances are defined in \eqref{ED_mV}. 
Further, if for the $i$-th SU the state of the PU at the sensing interval $d$ is denoted by $s_{i,d}\in\{0,1\}$, then for $\bdtheta_0\triangleq \{\mu_0,\sigma^2_0\}$ and $\bdtheta_1\triangleq \{\mu_1,\sigma^2_1\}$, the conditional 
probability distribution of $\bdx_i$ can be written as 
\begin{align}\label{x_distn}
 p(\bdx_i|\bds_i,\bdtheta_0,\bdtheta_1)
& =\prod^D_{d=1} \left(\N\left(x_{i,d}|\mu_1,\sigma^2_1\right)\right)^{\bdone
\left(s_{i,d}=1\right)}
\left(\N\left(x_{i,d}|\mu_0,\sigma^2_0\right)\right)^{\bdone\left(s_{i,d}=0\right)},
\end{align}
where $\bds_i=[s_{i,1},s_{i,2},\ldots,s_{i,D}]^T$ denotes the PU state vector, 
and $\bdone(A)$ is an indicator function which is one if $A$ is true, and 
is zero otherwise \cite{gubner2006}. Next, 
as discussed before and shown in Fig. \ref{fig:two_state}, we assume that 
the state vector $\bds_i$ follows a two-state Markov chain model with the transition probabilities $\alpha$ and $\beta$. 
Thus, the probability distribution of $\bds_i$ is written as 
\begin{align}\label{s_distn}
&p(\bds_i|\alpha,\beta) = p(s_{i,1})\prod^D_{d=2} p\left(s_{i,d}|s_{i,d-1}\right)\nonumber\\
&=  p(s_{i,1})\prod^D_{d=2} \left[\left(1-\alpha\right)^{\bdone\left(s_{i,d-1}=1\right)}
\beta^{\bdone\left(s_{i,d-1}=0\right)}\right]^{\bdone\left(s_{i,d}=1\right)}
\left[\alpha^{\bdone\left(s_{i,d-1}=1\right)}
\left(1-\beta\right)^{\bdone\left(s_{i,d-1}=0\right)}\right]^{\bdone\left(s_{i,d}=0\right)}.
\end{align}
where considering the 
steady-state distribution for the Markov process, we assume $s_{i,1}$ 
is Bernoulli distributed with mean $\frac{\beta}{\alpha+\beta}$.

Now if the model parameters of the above probability distributions are defined by 
$\bdTheta= \{\bdtheta_0,\bdtheta_1, \alpha,\beta\}$, 
an optimal scheme for estimating both $\bdTheta$ and $\bds_i$ for SU $i$ involves solving the following optimization problem 
\begin{align}
\left(\bds^*_i,\bdTheta^*\right)&=\argmax_{\left(
\bds_i,\bdTheta\right)} p(\bds_i,\bdTheta|\bdx_i)\nonumber\\
&=\argmax_{\left(
\bds_i,\bdTheta\right)} p\left(\bdx_i|\bds_i,\bdtheta_0,\bdtheta_1\right)
p\left(\bds_i|\alpha,\beta\right),
\end{align}
where for the sake of simplicity, we assumed a uniform prior distribution on $\bdTheta$. Note that 
due to the large dimensionality of the search space, jointly 
optimizing for $\bds_i$ and $\bdTheta$ is computationally difficult. 
Alternatively, we can aim to sequentially optimize for $\bds_i$ and $\bdTheta$ which 
involves solving the following two optimization problems: 
\begin{align}\label{opt_Theta_eqn}
\hat{\bdTheta} &= \argmax_{\bdTheta} \log p\left(\bdx_i|\bdTheta\right)\nonumber\\
&=\argmax_{\bdTheta}\log \sum_{\bds_i} p\left(\bdx_i, \bds_i|\bdTheta\right)
\end{align}
where \eqref{opt_Theta_eqn} maximizes the likelihood function of 
$\bdTheta$. Then using $\hat{\bdTheta}$ we can solve, 
\begin{align}\label{opt_si_eqn}
\hat{\bds}_i &=\argmax_{\bds_i} p\left(\bds_i|\bdx_i,\hat{\bdTheta}\right)\nonumber\\
&=\argmax_{\bds_i} p\left(\bdx_i|\bds_i,\hat{\bdtheta}_0,\hat{\bdtheta}_1\right)
p\left(\bds_i|\hat{\alpha},\hat{\beta}\right),
\end{align}

However, note that due to the log-sum in \eqref{opt_Theta_eqn}, 
directly optimizing for the elements of 
$\bdTheta$, e.g., using the derivative trick, 
does not result in the closed-form update 
equations, whereas using the numerical methods for optimization have the inherent 
complexity with the tuning of the step-size parameter \cite{boyd2004convex}. Furthermore, the optimization problem in \eqref{opt_si_eqn} is still a complex 
combinatorial search problem where the dimensionality of the 
search space increases exponentially with $D$. In the following, 
we develop a joint expectation maximization and Viterbi (EM-Viterbi) algorithm to estimate $\bds_i$ and 
$\bdTheta$ in a computationally efficient and optimal 
way using the closed-form update equations. 
\subsection{Expectation Maximization Algorithm}

An expectation maximization algorithm \cite{Dempster_EM, Bishop, 
Baum_1970} is an iterative algorithm which can be derived by first selecting a 
complete data model in order to compute an objective function of the model parameters.
Next, given an initial estimate of the 
parameters, it tends to improve this estimates in each iteration 
by maximizing the objective function which in turn maximizes the 
likelihood function\cite{Bishop}. The EM algorithm has 
been developed for a variety of estimation problems 
in recent years \cite{Rashid_TAES_2020, rashid2022EM, Bishop}, and in this subsection, we develop it to facilitate joint PU states 
and the model parameters estimation in order to enable distributed 
cooperative spectrum sensing using our proposed mWSED scheme.

To begin, let the complete data model for 
the $i$-th SU be denoted by $[\bdx^T_i, \bds^T_i]^T$, and suppose 
$\bdTheta^{(l-1)}$ is the $(l-1)$-st estimate of the model 
parameters, then in the $l$-th iteration it computes an expectation 
step (E-step) and a maximization step (M-step). In the E-step, 
it computes an expectation of the complete data log-likelihood 
function as follows 
\begin{equation}\label{Q_func}
\Q\left(\bdTheta;\bdTheta^{(l-1)}\right)=
\Exp_{p\left(\bds_i|\bdx_i,\bdTheta^{(l-1)}\right)}\left[\log 
p\left(\bdx_i,\bds_i|\bdTheta \right)\right],
\end{equation}
where we note that the expectation is with respect to the posterior 
distribution on $\bds_i$ given $\bdx_i$ and 
an old estimate $\bdTheta^{(l-1)}$. 
In the M-step, it maximizes the objective function in \eqref{Q_func} with respect to 
$\bdTheta$ by solving 
\begin{equation}\label{M_step_EM}
\bdTheta^{(l)} = \argmax_{\bdTheta} 
\Q \left(\bdTheta;\bdTheta^{(l-1)}\right),
\end{equation}
in which $\bdTheta^{(l)}$ represents the new estimate of $\bdTheta$ in 
the $l$-th iteration. The above E-step and M-step are repeated 
iteratively by replacing the old estimate with the new one 
until convergence is achieved. 
\begin{figure*}[ht]
\normalsize
\setcounter{mytempeqncnt}{\value{equation}}
\setcounter{equation}{16}

\begin{align}\label{Q_func_EM}
&\Q\left(\bdTheta;\bdTheta^{(l-1)}\right) =
\sum^D_{d=1}\left[\gamma(s_{i,d}=1)\log
\N\left(x_{i,d}|\mu_1,\sigma^2_1\right)+
\gamma(s_{i,d}=0)\log\N\left(x_{i,d}|\mu_0,\sigma^2_0\right)\right]+\nonumber\\
&\sum^D_{d=2}
\left[\xi(s_{i,d}=1,s_{i,d-1}=1)
\log(1-\alpha)+\xi(s_{i,d}=1,s_{i,d-1}=0)\log \beta+
\xi(s_{i,d}=0,s_{i,d-1}=1)\times\right.\nonumber\\
&\left.\log\alpha+ 
\xi(s_{i,d}=0,s_{i,d-1}=0)\log (1-\beta)
\right]+\text{const},
\end{align}

\setcounter{equation}{\value{mytempeqncnt}+1}
\hrulefill
\end{figure*}

Now using the distributions in \eqref{x_distn} and \eqref{s_distn}, and 
the following notation for the expectation operations, i.e., 
$\gamma(s_{i,d}=h)\triangleq\Exp[\bdone_{(s_{i,d}=h)}]$ and 
$\xi(s_{i,d}=h, s_{i,d-1}=g)\triangleq \Exp[\bdone_{(s_{i,d}=h)}\bdone_{(s_{i,d-1}=g)}]$, for $g,h\in\{0,1\}$, 
it can be easily derived that the objective 
function in \eqref{Q_func} can be written as shown in \eqref{Q_func_EM}. 
Further, note that $\gamma(s_{i,d}=h)= 
p\left(s_{i,d}=h|\bdx_i,\bdTheta^{(l-1)}\right)$ and 
$\xi(s_{i,d}=h, s_{i,d-1}=g)= 
p\left(s_{i,d}=h, s_{i,d-1}=g|\bdx_i,\bdTheta^{(l-1)}\right)$ 
for $g,h\in\{0,1\}$, where these probability distributions are 
derived in the Appendix. 

In order to compute the M-step in 
\eqref{M_step_EM}, we use the sequential optimization approach 
\cite{Rashid_TAES_2020, rashid2022EM} 
for simplicity, i.e., we maximize $\Q(\bdTheta;\bdTheta^{(l-1)})$ 
with respect to each parameter individually 
by keeping the others fixed to their current 
estimate. To that end, we use the derivative trick, and thus 
to maximize $\Q(\bdTheta;\bdTheta^{(l-1)})$ with respect to 
$\mu_h$ for $h\in\{0,1\}$, we compute its derivative and set 
it equal to zero as follows.
\begin{align}
\frac{\partial \Q(\bdTheta;\bdTheta^{(l-1)})}{\partial \mu_h} 
&=0\nonumber\\
\sum^D_{d=1} \left[
\gamma(s_{i,d}=h)\frac{(x_{i,d}-\mu_h)}{\sigma^2_h}\right]&=0,
\end{align}
solving it gives us a new estimate of $\mu_h$, in the $l$-th 
iteration of EM, which is written as\footnote{
{Note that upon EM's convergence, we can use \eqref{ED_mV} and 
thus divide the $\mu_0$'s estimate 
obtained from \eqref{mu_upd} 
by the number of samples per ED ($L$) to estimate 
the noise power of the PU's channel ($\sigma^2_n$). This noise power estimate can be 
used to compute the threshold using \eqref{ED_PdPfeqns} and \eqref{mWSED_PdPf_Eqns}.}}
\begin{align}\label{mu_upd}
\mu_h^{(l)}&=\frac{\sum^D_{d=1}\gamma(s_{i,d}=h)x_{i,d}}
{\sum^D_{d=1}\gamma(s_{i,d}=h)},
\end{align} 
for $h=0,1$. Now to maximize $\Q(\bdTheta;\bdTheta^{(l-1)})$ 
with respect to $\sigma^2_h$, we solve
\begin{align}
\frac{\partial \Q(\bdTheta;\bdTheta^{(l-1)})}{\partial \sigma^2_h} 
&=0\nonumber\\
\sum^D_{d=1} \left[
\gamma(s_{i,d}=h)\left(\frac{1}{2\sigma^2_h}-\frac{(x_{i,d}-\mu_h)^2}
{2\sigma^4_h}\right)\right]&=0,
\end{align}
from which we get the update equation for $\sigma^2_h$ as 
\begin{align}\label{sigma_upd}
\left(\sigma^2_h\right)^{(l)}&=
\frac{\sum^D_{d=1}\gamma(s_{i,d}=h)\left(x_{i,d}-\mu^{(l)}_h\right)^2}
{\sum^D_{d=1}\gamma(s_{i,d}=h)},
\end{align} 
where $h=0,1$. Similarly, using the same approach, it can 
be easily shown that the update equations for the 
Markov chain transition probabilities $\alpha$ and $\beta$ are 
given by
\begin{align}\label{alpha_upd}
\alpha^{(l)}
&=\frac{\sum^D_{d=2}\xi(s_{i,d}=0,s_{i,d-1}=1)}
{\sum^D_{d=2}[\xi(s_{i,d}=0,s_{i,d-1}=1)+\xi(s_{i,d}=1,s_{i,d-1}=1)]},\\
\beta^{(l)}
&=\frac{\sum^D_{d=2}\xi(s_{i,d}=1,s_{i,d-1}=0)}
{\sum^D_{d=2}[\xi(s_{i,d}=1,s_{i,d-1}=0)+\xi(s_{i,d}=0,s_{i,d-1}=0)]},
\label{beta_upd}
\end{align}
Thus, all the model parameters are updated iteratively in EM using 
the closed-form update Eqns. \eqref{mu_upd}, \eqref{sigma_upd}, 
\eqref{alpha_upd}, and \eqref{beta_upd} until convergence is achieved. 

Finally, upon the convergence of EM, the state vector 
$\bds_i=[s_{i,1},s_{i,2},\ldots,s_{i,D}]^T$ for the 
$i$-th user can be estimated, in a computationally efficient and optimal way, 
by using the Viterbi algorithm \cite{Viterbi_1973}. Thus, at SU $i$, 
let the EM estimate of the model parameters 
is denoted by $\hat{\bdTheta}$, then 
the Viterbi algorithm uses it to 
recursively solve the following optimization problem 
\begin{align}\label{est_s_eqn}
\omega_{i,d}(s_{i,d})
&=\max_{s_{i,d-1}}\left[p\left(x_{i,d}|s_{i,d},\hat{\bdTheta}\right)
p\left(s_{i,d}|s_{i,d-1},\hat{\bdTheta}\right)\omega_{i,d-1}(s_{i,d-1})\right],
\end{align}
for $d=2,3,\ldots,D$ with the initialization 
$\omega_{i,1}(s_{i,1})=p(x_{i,1}|s_{i,1},\hat{\bdTheta})p(s_{i,1},\hat{\bdTheta})$. The distributions 
$p\left(x_{i,d}|s_{i,d},\hat{\bdTheta}\right)$ and 
$p\left(s_{i,d}|s_{i,d-1},\hat{\bdTheta}\right)$ 
are given in \eqref{x_distn} 
and \eqref{s_distn}, respectively, and note that they are computed in \eqref{est_s_eqn} using only the required parameters estimate from 
the set $\hat{\bdTheta}$. Hence, by keeping track of the maximizing sequence at each time instant in \eqref{est_s_eqn} and by finding $\max_{s_{i,D}} \omega_{i,D}(s_{i,D})$ at time instant $D$, we can back trace the most probable sequence to get $\hat{\bds}_i$. 
\begin{algorithm}
  \footnotesize
\DontPrintSemicolon
\SetKwInput{KwPara}{Input}
\KwPara{$l=0$, $\bdx_i$ and $\bdx_j$ for $j\in N_i$, $\bdTheta^{(0)}$.}
\begin{enumerate}
\item[1)] Use the distributed consensus algorithm of \eqref{DCSS_eqn} 
to reach consensus \\on $\bdx_i$ with the other users in the network.
\end{enumerate}
\While{convergence criterion is not met} 
{
	$l=l+1$
\begin{enumerate}
\item[2)] Use the forward recursion in \eqref{nu_eqn} to compute 
$\nu_d(s_{i,d}=h)$ \\for all $d=1,2,\ldots,D$ and $h=0,1$.
\item[3)] Use the backward recursion in \eqref{pi_eqn} to compute 
$\pi_d(s_{i,d}=h)$ \\for all $d=D,D-1,\ldots,1$ and $h=0,1$.
\item[4)] Compute $\gamma(s_{i,d}=h)$ from \eqref{gamma_eqn} for all $d=1,2,\ldots,D$ \\and $h=0,1$, and compute $\xi(s_{i,d}=h,s_{i,d-1}=g)$ 
from \\ \eqref{xi_eqn} for all $d=2,\ldots,D$ and $h=0,1$, $g=0,1$.
\item[5)] Update the model parameters $\bdTheta^{(l)}$ 
using \eqref{mu_upd}, \eqref{sigma_upd}, \eqref{alpha_upd}, \\and \eqref{beta_upd}.
\end{enumerate}
}
\begin{enumerate}
\item[6)] Use the Viterbi algorithm in \eqref{est_s_eqn} to estimate the PU state vector $\hat{\bds}_i$ for 
SU $i$.
\item[7)] Compute the test statistics for mWSED using \eqref{mWSED_eqn} and 
compare it against a threshold to make PU detection. 
\end{enumerate}
\KwOut{$\hat{\bds}_i$, $\hat{\bdTheta}$, $T_i$}
\caption{States Estimation Based PU Detection for SU $i$ 
}\label{algo_1}
\end{algorithm} 
Note that the combination of EM and Viterbi algorithm is named here as 
the EM-Viterbi algorithm. 
However, once the state vector of the PU is estimated 
then we can use it in the mWSED 
algorithm proposed in Section \ref{mWSED_section} to 
combine only the highly correlated ED samples 
in its test statistic, and the resulting algorithm is referred to here as the 
EM-mWSED algorithm. Both EM-Viterbi and 
EM-mWSED are summarized for SU $i$ in Algorithm \ref{algo_1}. 
\begin{figure*}[t!]
    \begin{minipage}{0.33\textwidth}
     	\includegraphics[width=0.99\textwidth]{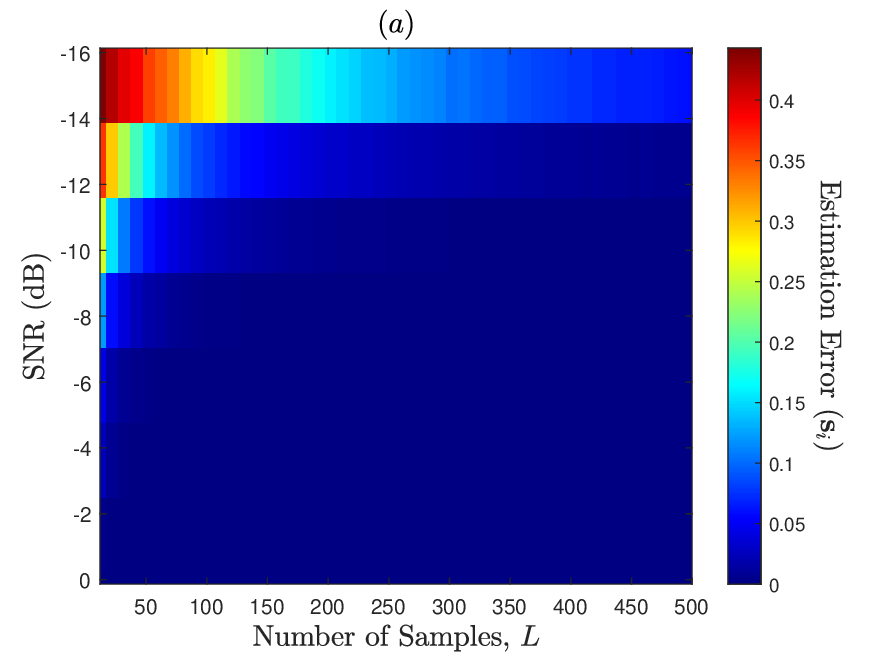}
    \end{minipage}\hspace{.01\linewidth}
    \begin{minipage}{0.33\textwidth}
\includegraphics[width=0.99\textwidth]{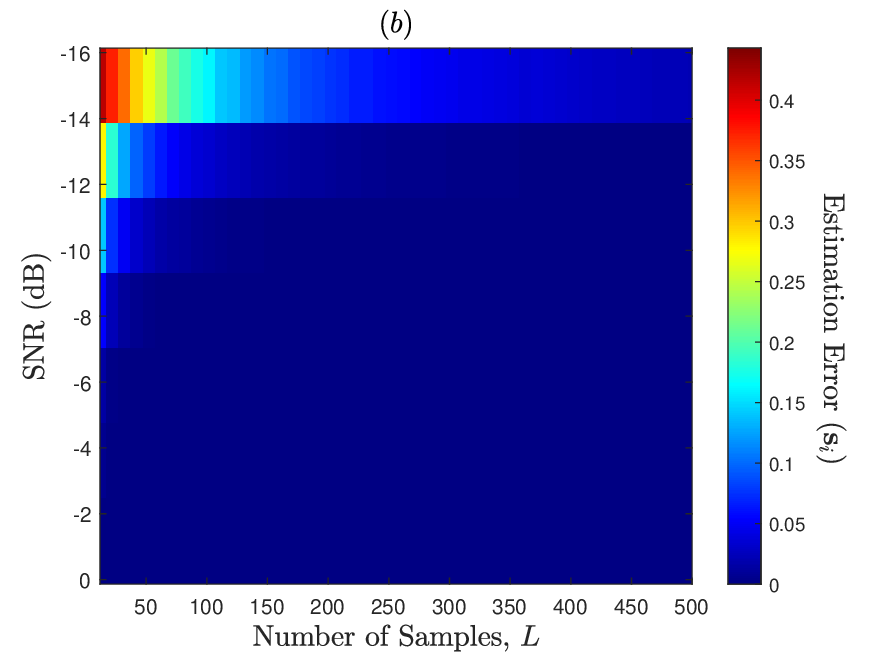}
		\end{minipage}
    \begin{minipage}{0.33\textwidth}
\includegraphics[width=0.99\textwidth]{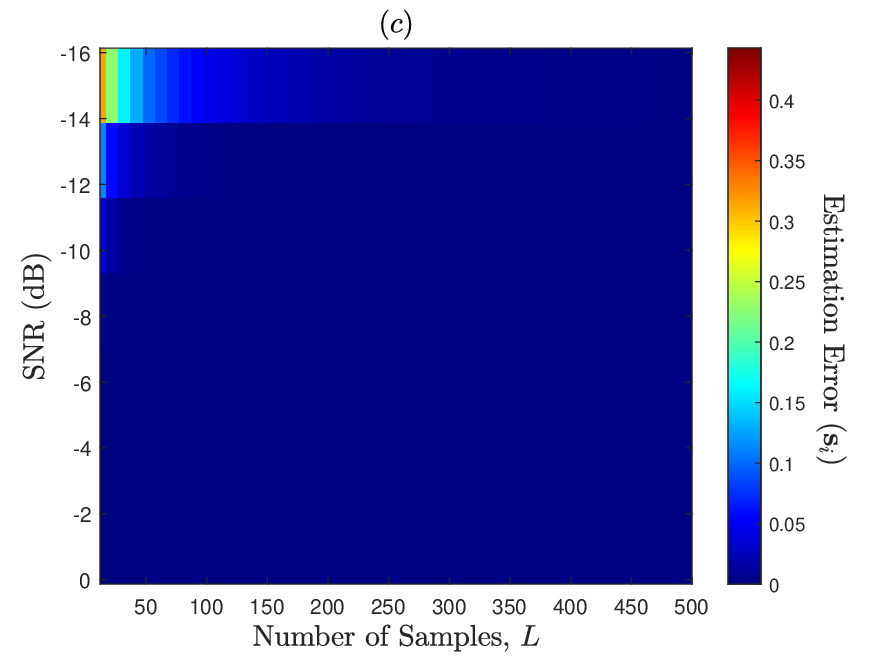}
		\end{minipage}
		\caption{Primary user states estimation error of EM-Viterbi for an SU 
		as a function of SNR (dB) and 
		the number of samples per energy statistics $L$, for $(a)$ $N=10$ SUs, 
		$(b)$ $N=20$ SUs, and $(c)$ $N=60$ SUs, and 
		when the network connectivity is $c=0.2$ and the 
		PU follows a highly dynamic profile with $\alpha=\beta=0.1$.}
				\label{fig_2:surf_err}
\end{figure*}

The computational complexity of EM-mWSED 
is dominated by the use 
of the distributed consensus algorithm of \eqref{DCSS_eqn} in 
Step $1$. This step has the complexity of $\mathcal{O}(|N_i|)$ 
per its iteration, where 
$|N_i|$ is the cardinality of the set of neighboring users of SU $i$. 
Furthermore, the forward and backward passes on the observation window 
in Steps $2$ and $3$, to compute 
the distributions in \eqref{nu_eqn} and \eqref{pi_eqn}, 
respectively, as well as 
the Viterbi algorithm in Step $6$ and \eqref{est_s_eqn} also dominate the computational complexity. 
These steps have the complexity of $\mathcal{O}(2D)$ 
where $D$ is the length of the observation window. Thus, 
the computational complexity of EM-mWSED is 
$\mathcal{O}(I_c|N_i|+2D I_e)$ where $I_c$ is the number of 
consensus iterations whereas $I_e$ denotes the number of EM iterations 
till convergence. 
The complexity of the energy detector based DCSS is 
$\mathcal{O}(I_c|N_i|)$ whereas that of the WSED based DCSS is 
$\mathcal{O}(I_c|N_i|+D)$. Thus, the performance improvement of 
EM-mWSED, as demonstrated in the next section, is at the cost of a  slight increase in the computational complexity per a single iteration of the EM algorithm.
\begin{figure}[tp]
        \centering
	\includegraphics[width=0.5\textwidth]{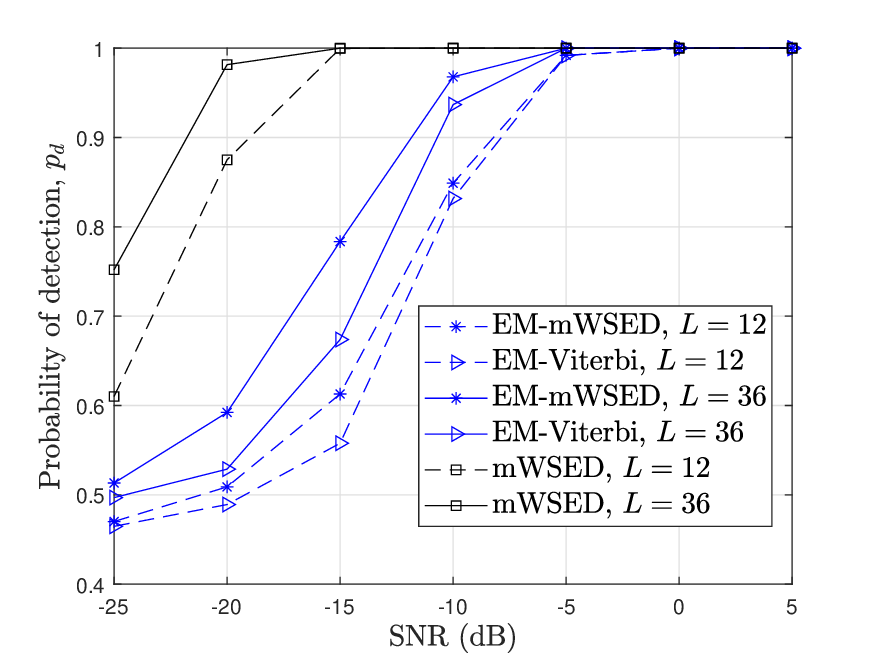}
	\caption{{Probability of detection vs. SNR of EM-mWSED, EM-Viterbi, and 
	mWSED for $N=20$ SUs and connectivity $c=0.5$, and when either $L=12$ or 
	$L=36$ samples are used per ED statistic. The PU follows a highly dynamic profile 
	with $\alpha=\beta=0.1$. }}
	\label{fig_3:EMViterbi_vs_EM-mWSED}
\end{figure}
\begin{figure}[tp]
        \centering
\includegraphics[width=0.5\textwidth]{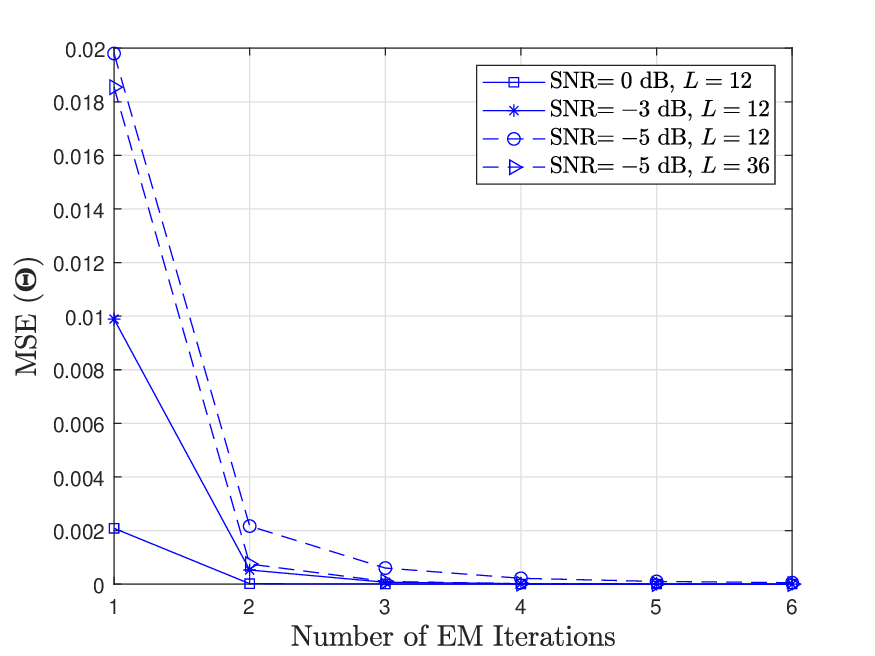}
	\caption{Mean-squared error (MSE) of estimating the model parameters $\bdTheta$ vs. EM iterations for $N=10$ SUs and connectivity $c=0.2$, 
	when SNR$=-5$ dB, $-3$ dB, or $0$ dB and 
		when the number of samples per energy statistics $L=12$ or $36$. The PU states 
		transition probabilities are $\alpha=\beta=0.1$.}
	\label{fig_3b:EM_convergence}
\end{figure}
\section{Simulation Results}\label{sim_results}

In this section, we demonstrate the {PU detection and 
states estimation performances of 
our proposed EM-Viterbi algorithm and the detection performance of our 
proposed EM-mWSED scheme.} 
For comparison purposes, 
we compare the performances of these algorithms 
to the conventional energy detector (ED), 
the weighted sequential energy detector (WSED) of \cite{Warit_2012, Warit_2014}, 
{multi-slot ED (msED) of \cite{KslotED_2022}, 
and the improved ED (IED) of \cite{IED_2012}} under 
different scenarios, when these detectors are 
used with the DCSS scheme and with the consensus happening 
on the present and past ED samples as proposed herein. As 
suggested in \cite{Warit_2012, Warit_2014} 
for WSED, we use a total of $3$ past 
ED samples in its test statistics for a highly 
dynamic PU, whereas ED uses only the present energy sample in its test statistics. 
{For the msED, we consider only the present and past ED samples 
for a fair comparison and plot its performance when $2$ slots are used in 
making the PU detection \cite{KslotED_2022}. Notably increasing the 
number of slots for msED results in a degraded performance for the 
considered time-varying PU.} We consider a network of $N$ secondary 
users randomly generated with a connectivity $c$ and the 
weighting matrix as defined in \eqref{DCSS_eqn}. The average 
number of connections per SU in the network is given by 
$R=c(N-1)$. The primary user follows a two-state Markov 
chain model to switch between the active and idle states 
with the transition probabilities $\alpha$ and $\beta$. The SUs 
collect $L$ samples individually 
to compute the energy statistic, and the length of the observation window is set 
as $D=150$. As described in Algorithm \ref{algo_1}, 
we assume that the SUs reach consensus on all the present and past ED samples over the 
observation window, which improves the SNR proportionally to 
${R}$ and aids in improving 
the detection and estimation performances. 
For the initialization of the EM algorithm, 
we determine the initial estimate of the means and variances by 
using the K-means clustering algorithm \cite{Bishop} over the window 
with $K=2$, whereas the initial estimate of the
transition probabilities can be computed by performing a coarse grid 
search over the likelihood function in \eqref{opt_Theta_eqn} in the 
$(0, 1)$ interval with a grid resolution of $0.1$. 

In Fig. \ref{fig_2:surf_err}, we demonstrate the PU's states estimation performance 
of the EM-Viterbi algorithm as a function of SNR (dB) and the number of samples 
per energy statistics ($L$). The states estimation error for SU $i$ 
is defined here as $\text{Estimation Error }(\bds_i)=\frac{1}{D}\sum^D_{d=1}
\Exp\left[\bdone\left(\hat{s}_{i,d}\neq s_{i,d}\right)\right]$ where 
the expectation is computed over several Monte Carlo trials. 
Further, we assume that the consensus is reached on the ED samples in 
Step $1$ of Algorithm \ref{algo_1} prior to estimation, and thus the error 
plots in this figure are observed at 
all the SUs in the network. We consider here that the 
secondary users network has $N=10$, $20$, and $60$ users with 
connectivity $c=0.2$. The PU displays a highly dynamic nature with 
transition probabilities $\alpha=\beta=0.1$. 
{Firstly, it is observed} that 
for a fewer number of SUs ($N=10$) in the network, the estimation error 
is higher at the lower SNRs and the lower $L$ values, but when the 
number of SUs in the network increases from $N=10$ to $20$ and then 
to $60$, the estimation error decreases significantly even 
for the lower SNR and $L$ values. This is because 
the SNR upon consensus in Step 1 of Algorithm \ref{algo_1} improves 
proportionally to ${R}$, as each SU exploits about $R$ independent 
ED samples of the PU's channel. 
{This implies that better performance can be achiever at lower PU's SNR 
environments by deploying DCSS-based 
larger and highly connected networks of SUs, i.e., networks 
with larger $N$ and $c$ values.}, 
This explains our motivation behind using DCSS scheme prior to the estimation process in Algorithm \ref{algo_1}. 
{Secondly, as discussed in Section \ref{Intro_section}}, it is observed that the estimation error is higher at lower SNR and $L$ values for all the considered cases in Fig. \ref{fig_2:surf_err}. {As discussed in Section \ref{sys_mod}, 
this is because the distribution of the ED samples under 
the two hypotheses highly 
overlap at those values which degrades the 
estimation and detection performances, but increasing $L$ at lower SNRs 
separates the means of 
the two distributions that in turn improves the performance\footref{ED_footnote}.
However, larger $L$ implies larger sensing durations which 
reduces the network throughput and increases the energy consumption \cite{Liu_2020}.}
\begin{figure}[tp]
        \centering
\includegraphics[width=0.5\textwidth]{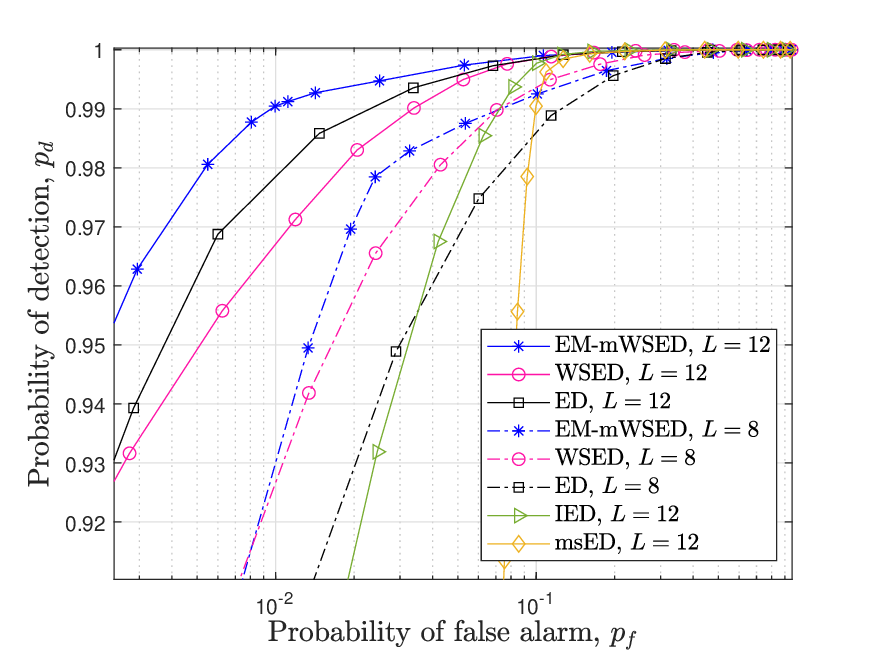}
			\caption{{Receiver operating characteristic curves of EM-mWSED, 
			conventional ED, WSED, IED, and msED for different 
			number of samples per energy statistics $L$, when $N=20$ SUs, 
	connectivity $c=0.2$, SNR $=-5$ dB, 
	and the PU states transition probabilities $\alpha=\beta=0.1$ 
	are used.}}
			\label{fig_5:EM_ROC1}
\end{figure}
\begin{figure}[tp]
        \centering
\includegraphics[width=0.5\textwidth]{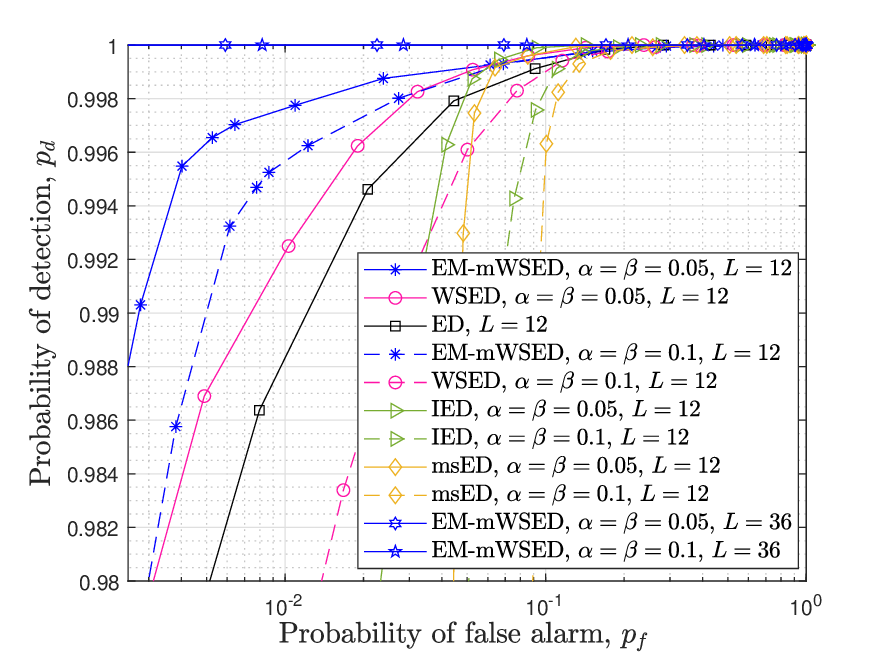}
	\caption{{Receiver operating characteristic curves of EM-mWSED, 
	conventional ED, WSED, IED, and msED for slowly time-varying PU with 
			$\alpha=\beta=0.05$ and a highly dynamic PU with $\alpha=\beta=0.1$. 
	The number of SUs $N=10$, 
	connectivity $c=0.2$, SNR$=-3$ dB, and $L=12$ or $36$.}}
	\label{fig_6:EM_ROC2}		
\end{figure}

{In general, a poor estimation error for PU's states do effect 
the detection performances of EM-Viterbi\footnote{Since 
the last element $\hat{s}_{i,D}$ in the state vector estimate, $\hat{\bds}_i$ 
at SU $i$, represents 
an estimate of the PU's state in the present sensing period $D$, EM-Viterbi can 
also be used for PU detection.} and EM-mWSED schemes, 
but for an appropriate choice of $L$ 
value and by aggregating present and past ED samples, we 
can improve the detection performance at lower SNRs. To illustrate this fact, 
in Fig. \ref{fig_3:EMViterbi_vs_EM-mWSED}, we compare the detection 
performances of EM-Viterbi and EM-mWSED by varying SNR and $L$ values, when 
$N=20$, $c=0.5$, and $\alpha=\beta=0.1$. Note 
that the state vector estimated by EM-Viterbi is used in EM-mWSED to 
average the highly correlated present and past ED samples. 
The weights on the aggregated ED samples in EM-mWSED are chosen as exponentially distributed as it results in better detection 
performance than equal weighting at lower SNR or $L$ values, due to the 
rise in the estimation error. 
Further, the threshold at each SNR and $L$ values 
is selected based on the false alarm rate of EM-Viterbi. 
However, it is observed that the 
detection performance of EM-mWSED is better than that of EM-Viterbi at 
lower SNRs, and that the performance of EM-mWSED improves significantly by a 
slight increase in the $L$ value.  While the increase in $L$ 
aids in improving the EM's estimation performance, 
the better performance of mWSED is due to 
the fact that it averages the highly correlated present and past 
ED samples in its test statistics which reduces the variances under the two hypotheses as illustrated in Section \ref{sim_mWSED}. This improves the detection 
performance at lower SNRs and smaller $L$ values. 
Finally, as expected, we also observe that the performances of both 
algorithms achieve that of mWSED with an increase in SNR or $L$ values.} 

In Fig. \ref{fig_3b:EM_convergence}, we illustrate the convergence 
performance of the EM algorithm in estimating the model parameters 
$\bdTheta$ when $N=10$ SUs are considered in the network with 
network connectivity $c=0.2$. 
The mean-squared error (MSE) of $\bdTheta$ is 
defined as $\text{MSE}\left(\bdTheta\right)=\Exp\left[\mid\mid\bdTheta-\hat{\bdTheta}\mid\mid_2\right]$. It is observed that for $L=12$ 
samples per energy statistics, as the SNR increase from $-5$ dB, to 
$-3$ dB, and then to $0$ dB, the EM algorithm converges faster in fewer 
iterations. 
Similar observation is also made when for a lower SNR value of $-5$ dB, we increase $L$ from $12$ to $36$. This is due to the fact that the initial estimates for EM are improved at the larger SNR and $L$ values which results in its faster convergence response and improved estimation performance.

While EM-Viterbi provides a single operating point for SUs due 
to the estimation of the present state, in terms of detection 
probability and false alarm probability; in contrast, 
by using the estimated state vector 
in mWSED, the EM-mWSED algorithm can provide a wide range of operating points for SUs.
{Further, as observed in Fig. \ref{fig_3:EMViterbi_vs_EM-mWSED}, 
EM-mWSED outperforms EM-Viterbi at lower SNRs which makes it a preferable 
choice. As such, in Figs. \ref{fig_5:EM_ROC1} and \ref{fig_6:EM_ROC2}, we show the receiver operating characteristic (ROC) curves of the proposed EM-mWSED algorithm and compare it with those of the ED, WSED, IED, and msED schemes.} 
In Fig. \ref{fig_5:EM_ROC1}, 
we consider $N=20$ SUs in the network with connectivity $c=0.2$. The PU states transitioning probabilities 
are selected as $\alpha=\beta=0.1$. The SNR is assumed to be 
$-5$ dB, whereas the number of samples per energy statistic $L$ is 
assumed to be either $8$ or $12$. 
As expected, it is observed that EM-mWSED outperforms {all other methods} 
at increasing the detection probability and reducing the false alarm probability, 
and thereby provides a wide range of operating points for SU. Furthermore, 
the detection performance of EM-mWSED improves with the increase in the 
$L$ values, due to decrease in the states estimation error.

Fig. \ref{fig_6:EM_ROC2} compares the ROC curves of 
EM-mWSED with that of {ED, WSED, IED, and msED} schemes. 
The PU is either considered 
to be slowly time-varying with $\alpha=\beta=0.05$ or highly dynamic 
with $\alpha=\beta=0.1$ as considered earlier. There are 
$N=10$ SUs in the network with connectivity $c=0.2$, and the SNR is considered 
to be $-3$ dB with the number of samples per energy statistics as either 
$L=12$ or $36$. Firstly, by comparing Figs. \ref{fig_5:EM_ROC1} and 
\ref{fig_6:EM_ROC2} for $L=12$, SNR$=-3$ dB, and $\alpha=\beta=0.1$, 
we observe a decay in 
the detection performance of EM-mWSED due to decrease in the value of 
$R$ in Fig. \ref{fig_6:EM_ROC2}, 
which reduces the SNR upon consensus as discussed above. 
Secondly, it is observed in Fig. \ref{fig_6:EM_ROC2} 
that for both slowly and highly dynamic natures of the PU, 
our EM-mWSED algorithm performs better than the other detectors 
as expected. However, when $L=12$, its detection performance appears to 
be dependent on the time-varying nature of the PU, and 
it is seen to be better in case of slowly time-varying PU than a highly dynamic PU. This is because at the lower SNR or $L$ values, EM-Viterbi 
can easily characterize the ED samples, corresponding to the two states of PU, 
when the PU is slowly time-varying than when it is highly dynamic, 
and thus results in a lower estimation error in the former case. 
However, when $L$ increases to $36$, then the estimation error of EM-Viterbi 
decreases for a highly dynamic PU as well, 
which in turn results in the similar performance of EM-mWSED 
for both kinds of PUs, as shown in this figure. 

Finally, while the focus herein is on investigating the detection vs. false alarm probabilities, it is worth noting that, on the one hand, where the throughput 
performance of an SU 
under $H_1$ is proportional to ${(1-p_d)}$, on the other hand, the 
throughput under $H_0$ is proportional to $(1-p_f)$ \cite{Rozeha_2011}. 
Thus, the higher 
detection probability of EM-mWSED as compared with that of ED, WSED, IED, and msED  implies a higher throughput of SUs with reduced interference to the primary user, whereas its capability of simultaneously decreasing the 
false alarm probability with the increase in the average connections 
per SUs in the network, SNR, or $L$, 
implies a higher throughput during the idle state of the PU.

\section{Conclusion}\label{conclusion_section}

We considered the problem of DCSS {at lower SNRs} when the present and 
past ED samples are aggregated in a test statistic for improved 
PU detection {with fewer samples 
per energy statistics. Using a few samples per energy aids 
in reducing the sensing duration and power consumption per SU. Furthermore, 
it increases the throughput of the network}. 
A modified weighted 
sequential energy detector is proposed which 
utilizes the PU states information over an observation window to combine 
only the highly correlated ED samples in its test statistics. 
In practice, the states information is unknown, and thus we developed 
an EM-Viterbi algorithm to iteratively estimate them using the 
ED samples collected over the window. The estimated states are then 
used in mWSED to compute its test statistics, and the resulting algorithm 
is named here as the EM-mWSED algorithm. Simulation results are included 
to demonstrate the estimation performance of EM-Viterbi and compare the 
detection performance of both EM-Viterbi and EM-mWSED {schemes. 
Furthermore, the detection performance of EM-mWSED is compared with that of the 
conventional ED, WSED, IED, and msED methods. The results demonstrate 
that our proposed EM-mWSED performs better than the other schemes}, 
and its performance improves by either increasing the average 
number of connections per SU in the network, or by increasing the 
SNR or the number of samples per energy statistics, for both slowly varying 
and highly dynamic PU. 

\appendix

Herein, we present the derivation of the 
probabilities $\gamma(s_{i,d}=h)= 
p\left(s_{i,d}=h|\bdx_i,\bdTheta^{(l-1)}\right)$ and 
$\xi(s_{i,d}=h, s_{i,d-1}=g)= 
p\left(s_{i,d}=h, s_{i,d-1}=g|\bdx_i,\bdTheta^{(l-1)}\right)$ 
for $g,h\in\{0,1\}$ and the $l$-th iteration of EM.
To begin, the posterior distribution of $s_{i,d}$ given 
$\bdx_i$ and $\bdTheta^{(l-1)}$ can be written as 
\begin{align}
&p\left(s_{i,d}|\bdx_i,\bdTheta^{(l-1)}\right)\nonumber\\
&\propto 
p\left(s_{i,d},\bdx_i|\bdTheta^{(l-1)}\right)\nonumber\\
&=p\left(s_{i,d},\bdx_{i,1:d},\bdx_{i,d+1:D}|\bdTheta^{(l-1)}\right)\nonumber\\
&=p\left(\bdx_{i,d+1:D}|s_{i,d},\bdTheta^{(l-1)}\right)
p\left(\bdx_{i,1:d},s_{i,d}|\bdTheta^{(l-1)}\right)\nonumber\\
&\triangleq \pi_d(s_{i,d})\nu_d(s_{i,d}),
\end{align} 
where the distributions $\pi_d(s_{i,d})$ and $\nu_d(s_{i,d})$ 
are computed later herein. The notation $\bdx_{i,m:n}\triangleq 
[x_{i,m},x_{i,m+1},\ldots,x_{i,n}]^T$ which is a shorthand to 
represent the elements in $\bdx_i$ from index $m$ to $n$ where 
$m,n\in\{1,2,\ldots,D\}$. Similarly, we can write the joint 
distribution of $s_{i,d}$ and $s_{i,d-1}$ as 
\begin{align}
&p\left(s_{i,d}, s_{i,d-1}|\bdx_i,\bdTheta^{(l-1)}\right)\nonumber\\
&\propto p\left(\bdx_{i,1:d-1},s_{i,d-1}, s_{i,d},\bdx_{i,d:D}, 
\bdTheta^{(l-1)}\right)\nonumber\\
&=p\left(\bdx_{i,1:d-1},s_{i,d-1}|\bdTheta^{(l-1)}\right)
p\left(\bdx_{i,d:D},s_{i,d}|s_{i,d-1},\bdTheta^{(l-1)}\right)\nonumber\\
&=\nu_{d-1}(s_{i,d-1})p\left(x_{i,d},\bdx_{i,d+1:D},s_{i,d}|s_{i,d-1},\bdTheta^{(l-1)}\right)\nonumber\\
&=\nu_{d-1}(s_{i,d-1})\pi_d(s_{i,d})p\left(x_{i,d}|s_{i,d},\bdTheta^{(l-1)}\right)
p\left(s_{i,d}|s_{i,d-1},\bdTheta^{(l-1)}\right),
\end{align}
where the conditional distribution of $x_{i,d}$ and the conditional 
prior distribution of $s_{i,d}$ that are used above are both defined in \eqref{x_distn} and \eqref{s_distn}. Next we follow the forward-backward recursion approach in \cite{Baum_1970} to compute the distributions $\nu_d(s_{i,d})$ and $\pi_d(s_{i,d})$. 
First, to compute $\nu_d(s_{i,d})$, we write 
\begin{align}\label{nu_eqn}
&\nu_d(s_{i,d})=p\left(s_{i,d},\bdx_{i,1:d}|\bdTheta^{(l-1)}\right)\nonumber\\
&=\sum_{s_{i,d-1}}p\left(s_{i,d},s_{i,d-1},\bdx_{i,1:d-1},x_{i,d}|\bdTheta^{(l-1)}\right)\nonumber\\
&=\sum_{s_{i,d-1}}p\left(x_{i,d}|s_{i,d},\bdTheta^{(l-1)}\right)
p\left(s_{i,d}|s_{i,d-1},\bdTheta^{(l-1)}\right)
p\left(s_{i,d-1},\bdx_{i,d-1}|\bdTheta^{(l-1)}\right)\nonumber\\
&=\sum_{s_{i,d-1}}c(s_{i,d},s_{i,d-1}) \nu_{d-1}(s_{i,d-1}),
\end{align}
where we have defined  
$c(s_{i,d},s_{i,d-1})=p\left(x_{i,d}|s_{i,d},\bdTheta^{(l-1)}\right)$ $p\left(s_{i,d}|s_{i,d-1},\bdTheta^{(l-1)}\right)$, and in \eqref{nu_eqn} the summation is over $s_{i,d-1}\in\{0,1\}$. The forward recursion in \eqref{nu_eqn} occurs in the $l$-iteration of EM 
for all $d=2,3,\ldots,D$ with the initialization 
$\nu_1(s_{i,1})=p(s_{i,1}|\bdTheta^{(l-1)})
p(x_{i,1}|s_{i,1},\bdTheta^{(l-1)})$ which is defined in \eqref{x_distn} and \eqref{s_distn}. Next we write the backward 
recursion equation to compute $\pi_d(s_{i,d})$ as follows
\begin{align}\label{pi_eqn}
&\pi_d(s_{i,d})=p\left(\bdx_{i,d+1:D}|s_{i,d},\bdTheta^{(l)}\right)\nonumber\\
&=\sum_{s_{i,d+1}}p\left(\bdx_{i,d+1:D},s_{i,d+1}|s_{i,d},\bdTheta^{(l)}\right)\nonumber\\
&=\sum_{s_{i,d+1}} p\left(\bdx_{i,d+2:D}|s_{i,d+1},\bdTheta^{(l)}\right)
p\left(x_{i,d+1}|s_{i,d+1},\bdTheta^{(l)}\right)
p\left(s_{i,d+1}|s_{i,d},\bdTheta^{(l)}\right)\nonumber\\
&=\sum_{s_{i,d+1}} \pi_{d+1}(s_{i,d+1})
c(s_{i,d+1},s_{i,d}),
\end{align}
in which the summation runs over $s_{i,d+1}\in\{0,1\}$ for all 
$d=D-1,D-2,\ldots,1$ with the initialization $\pi_D(s_{i,D})=1$. 
Finally, the probabilities $\gamma(s_{i,d}=h)= 
p\left(s_{i,d}=h|\bdx_i,\bdTheta^{(l-1)}\right)$ and 
$\xi(s_{i,d}=h, s_{i,d-1}=g)= 
p\left(s_{i,d}=h, s_{i,d-1}=g|\bdx_i,\bdTheta^{(l-1)}\right)$ 
for $g,h\in\{0,1\}$ can be computed as 
\begin{align}\label{gamma_eqn}
\gamma(s_{i,d}=h)&=p\left(s_{i,d}=h|\bdx_i,\bdTheta^{(l-1)}\right)\nonumber\\
&=\frac{\nu_d(s_{i,d}=h)\pi_d(s_{i,d}=h)}{\sum_{s_{i,d}}\nu_d(s_{i,d})\pi_d(s_{i,d})},
\end{align}
and,
\begin{align}\label{xi_eqn}
&\xi(s_{i,d}=h, s_{i,d-1}=g)\nonumber\\
&= p\left(s_{i,d}=h, s_{i,d-1}=g|\bdx_i,\bdTheta^{(l-1)}\right)\nonumber\\
&=\frac{\nu_{d-1}(s_{i,d-1}=g)\pi_d(s_{i,d}=h)c(s_{i,d}=h,s_{i,d-1}=g)}{\sum_{s_{i,d}}\sum_{s_{i,d-1}}\nu_{d-1}(s_{i,d-1})\pi_d(s_{i,d})c(s_{i,d},s_{i,d-1})},
\end{align}
\bibliographystyle{IEEEtran}
\bibliography{References}
%

\end{document}